\documentclass[fleqn,usenatbib,useAMS]{mnras}

\usepackage[percent]{overpic}
\usepackage{newtxtext,newtxmath}

\usepackage[T1]{fontenc}
\usepackage{ae,aecompl}
\usepackage{multirow}

\usepackage{graphicx}	
\usepackage{amsmath}	

\title[FRB source evolution]{The fast radio burst population evolves, consistent with the star-formation rate}

\author[C.W.~James et al.]{
C.W.~James,${^{1}}$\thanks{E-mail: clancy.james@curtin.edu.au}
J.X.~Prochaska${^{2,3}}$
J.-P.~Macquart,${{^1}}$
F.O.~North-Hickey${^{1}}$
K.~W.~Bannister${^{4}}$\newauthor
 and
A.~Dunning${^{4}}$
\\
$^{1}$International Centre for Radio Astronomy Research, Curtin University, Bentley, WA 6102, Australia \\
$^{2}$Kavli Institute for the Physics and Mathematics of the Universe, 5-1-5 Kashiwanoha, Kashiwa 277-8583, Japan.\\
$^{3}$Astronomy Department, University of Washington, Seattle, WA 98195, USA.\\
$^4$CSIRO Astronomy and Space Science, PO Box 76, Epping, NSW 1710, Australia.
}

\date{Accepted XXX. Received YYY; in original form ZZZ}

\pubyear{2020}

\begin{document}

\label{firstpage}
\pagerange{\pageref{firstpage}--\pageref{lastpage}}
\maketitle

\begin{abstract}
Fast radio bursts (FRBs) are extremely powerful sources of radio waves observed at cosmological distances. We use a sophisticated model of FRB observations --- presented in detail in a companion paper --- to fit FRB population parameters using large samples of FRBs detected by ASKAP and Parkes, including seven sources with confirmed host galaxies.
Our fitted parameters demonstrate that the FRB population evolves with redshift in a manner consistent with, or faster than, the star-formation rate (SFR), ruling out a non-evolving population at 99.9\% C.L. Our estimated maximum FRB energy is $\log_{10} E_{\rm max} [{\rm erg}] = 41.84_{-0.18}^{+0.49}$ (68\% C.L.) assuming a 1\,GHz emission bandwidth, with slope of the cumulative luminosity distribution $\gamma=-1.16_{-0.12}^{+0.11}$. 
We find a log-mean host DM contribution of $145_{-60}^{+64}$\,pc\,cm$^{-3}$ on top of a typical local (ISM and halo) contribution of $\sim80$\,pc\,cm$^{-3}$, which is higher than most literature values. These results are consistent with the model of FRBs arising as the high-energy limit of magnetar bursts, but allow for FRB progenitors that evolve faster than the SFR.
\end{abstract}

\begin{keywords}
radio continuum: transients -- methods: statistical
\end{keywords}



\section{Introduction}

Fast radio bursts (FRBs) are extragalactic transient radio sources of millisecond duration \citep{Lorimer2007,Thornton2013}. Some repeat, while most have not been observed to do so \citep{Spitler2016,CHIME2019b,CHIME2019c,CHIME2020a,Kumar2019,Shannonetal2018,James2020a}, and the question of whether or not there are one, two, or more FRB populations remains open. The recent observation of a Galactic magnetar flare with FRB-like properties strongly suggests such objects as an FRB progenitor class \citep{INTEGRALMagnetar2020,Magnetar_CHIME,Magnetar_STARE2}, though many more have been proposed \citep{Platts2018}. Yet, this flare was three orders of magnitude less powerful than the weakest FRBs, which in turn are orders of magnitude weaker than the most powerful FRBs \citep{Shannonetal2018}. FRBs may therefore have an unrelated origin.

If the FRB population does originate from young magnetars, they would be expected to be closely correlated with star-forming activities, as observed for two rapid repeaters \citep{Tendulkar2017,MarcoteRepeaterLocalisation2020}. However, the single largest sample of localised FRBs comes from the Australian Square Kilometre Array Pathfinder \citep[ASKAP; ][]{Bannister2019,Prochaska2019,Bhandari2020b}. The host galaxies of these FRBs --- which due to ASKAP's large field of view (FOV) and higher detection threshold tend to be the intrinsically most powerful bursts --- do not show evidence for unusual star-forming activity \citep{Bhandari2020a,Heintz2020}. This allows for the possibility of much of this population to arise from other sources, e.g.\ compact binary mergers \citep[see ][and references contained therein]{Caleb2018}.


A useful method to distinguish between these models comes from the evolution of the FRB population on cosmological timescales. If FRBs originate from young magnetars, they will closely follow the star-formation rate (SFR) \citep{Metzger2017magnetar}, and peak in the redshift range 1--3. A binary merger scenario however would likely lag the SFR, and possibly result in an FRB rate that is increasing with cosmological time \citep{Cao2018delayedmergers}.
As yet, FRB population analysis has not been able to distinguish between these scenarios \citep{Arcus2020,Luo2020}. Other methods yield mixed results: \citet{Hashimoto2020evolution} find evidence against the redshift evolution of once-off FRBs, and some evidence for redshift evolution of the event rate for repeating FRBs; while \citet{Locatelli2019} find evidence for an evolving FRB population for once-off FRBs. However, neither work follows a comprehensive approach advocated by \citet{Connor2019}, by modelling observational biases, and allowing for the confounding effects of the FRB luminosity function.

The FRB luminosity function is interesting in and of itself.
Comparisons of the luminosity function of individual repeaters \citep[e.g.][]{Law2017} to the population as a whole tests the credibility of the one-population model, while evidence for a minimum burst energy above that produced by Galactic magnetars would require a separate progenitor class, or at least a separate emission mechanism. Models requiring rare events to explain FRBs can be challenged by measurements of the absolute volumetric rate \citep{Ravi2019}. Estimates of the maximum FRB energy not only challenges theoretical models and pushes up against theoretical limits \citep{Lu2018radiation}, but affects the ability to use FRBs as cosmological probes.
Estimates of the host contribution to dispersion measure (DM) inform us of the environment surrounding FRB progenitors.
Consequently, several groups have begun modelling the FRB population in an attempt to derive these parameters, although the results and methods have been inconsistent \citep{Calebetal2016,Luo2018,Lu2019,Luo2020,Arcus2020,Gardenier2020FRBpoppy}.

In a companion paper \citep{James2021Full}, we present our method to model the FRB population. It uses the the methodology advocated by \citet{Connor2019}, and first implemented by \citet{Luo2020}, while making several significant advances in accuracy and precision, and taking advantage of recent FRB localisations, and fitting for the measured signal-to-noise ratio.
This models all known observational biases in detail, allowing us to make accurate and precise estimates of FRB population parameters, and model its cosmological source evolution. Here, we present maximum-likelihood estimates of FRB population parameters using FRBs observed by the Australian Square Kilometre Array Pathfinder (ASKAP) and Parkes, and discuss the implications for the FRB population.

\section{Review of the model}
\label{sec:modelling}

In modelling FRB observations, it is critically important to account for a range of observational biases. Our full treatment is contained in a (much lengthier) companion paper, 
\citet{James2021Full}. To briefly summarise, we account for telescope beamshape, and reduced observational sensitivity to high-DM, high-width FRBs, as recommended by \citet{Connor2019}; and fluctuations in cosmological dispersion measure according to best-fit cosmological parameters, local contributions from the Milky Way's interstellar medium (ISM) and halo, and a log-normal distribution $p({\rm DM}_{\rm host}^\prime)$ of the host DM contribution, as per \citet{Macquart2020}. This latter contribution, defined by
\begin{eqnarray}
p({\rm DM}_{\rm host}^\prime) = \frac{1}{{\rm DM_{\rm host}^\prime}} \frac{1}{\sigma_{\rm host} \sqrt{2 \pi}} 
e^{-\frac{(\log {\rm DM}^\prime_{\rm host}-\mu_{\rm host})^2}{2 \sigma_{\rm host}^2}}, \label{eq:phost}
\end{eqnarray}
is fit using the parameters $\mu_{\rm host}$ and $\sigma_{\rm host}$. The effective host DM, DM$_{\rm host}$, corrects the host DM for redshift: DM$_{\rm host} = {\rm DM}_{\rm host}^\prime/(1+z)$.

Our model for the FRB population uses a power-law with cumulative slope $\gamma$ and maximum energy $E_{\rm max}$, such that the probability of observing an FRB above an energy threshold $E_{\rm th}$ is given by
\begin{eqnarray}
p(E>E_{\rm th}) & = & \frac{ \left( \frac{E_{\rm th}}{E_{\rm min}}\right)^\gamma - \left(\frac{E_{\rm max}}{E_{\rm min}}\right)^\gamma  }{1-\left(\frac{E_{\rm max}}{E_{\rm min}}\right)^\gamma}. \label{eq:integral_energy_distribution}
\end{eqnarray}
The minimum FRB energy is not well-constrained by current observations, and is set to a very low value of $10^{30}$\,erg. We scale the FRB energy $E$ according to $E \sim \nu^\alpha$; for data taken exclusively at L-band ($\sim 1.4$\,GHz), the model is almost degenerate to $\alpha$ \citep[a conclusion also reached by][]{Lu2019,Arcus2020}, and so we use a symmetric Gaussian prior of $\alpha=-1.5 \pm 0.3$ \citep{Macquart2019a}.

We model the evolution of the FRB population $\Phi(z)$ (bursts per proper time per comoving volume) by smoothly scaling the SFR with the parameter $n$,
\begin{eqnarray}
\Phi(z) & = & \frac{\Phi_0}{1+z} \left( \frac{{\rm SFR}(z)}{{\rm SFR}(0)} \right)^n. \label{eq:phiz}
\end{eqnarray}
We take SFR$(z)$ from \citet{MadauDickinson_SFR},
\begin{eqnarray}
{\rm SFR}(z) & = & 1.0025738 \frac{(1+z)^{2.7}}{(1 + \left(\frac{1+z}{2.9}\right)^{5.6}}. \label{eq:sfr_n}
\end{eqnarray}
Thus our full model treats $E_{\rm max}$, $\gamma$, $\alpha$, $n$,  $\mu_{\rm host}$ and $\sigma_{\rm host}$ as free parameters.

There is some ambiguity in the interpretation of $\alpha$, which can instead be interpreted as a frequency-dependent rate. This is motivated by the many FRBs with a limited band occupancy, as originally noted by \citet{Law2017} for FRB~121102. The interpretation of $\alpha$ has slight effects on the modelling: in the rate interpretation, the FRB rate at high $z$ is modified directly through a further factor $\Phi(z) \sim (1+z)^{\alpha}$, while in the spectral-index interpretation, this occurs indirectly, through the k-correction affecting the calculation of threshold energy $E_{\rm th}$ from a fluence threshold $F_{\rm th}$, and hence the rate via Eq.~\ref{eq:integral_energy_distribution}. In the absence of an obvious correct treatment, we by default present results for the spectral index interpretation, and show results for the rate interpretation in Appendix~\ref{sec:extra_data}.

We use a sample of 24 non-localised, and seven localised, FRBs detected by ASKAP, and 20 FRBs detected by the Parkes multibeam system. These have been selected due to them occurring at high Galactic latitudes where the reduced sensitivity due to high Galactic DM is unimportant. The full telescope beamshape of each of these instruments is modelled in detail in our companion paper, based off the methods of \citet{James2019cSensitivity}, while the reduction in sensitivity to high DMs and widths is modelled using the time- and frequency resolutions of the instruments according to \citet{Cordes_McLaughlin_2003}.

\section{Results}
\label{sec:results1}

\begin{figure*}
    \centering
    \includegraphics[width=0.32\textwidth]{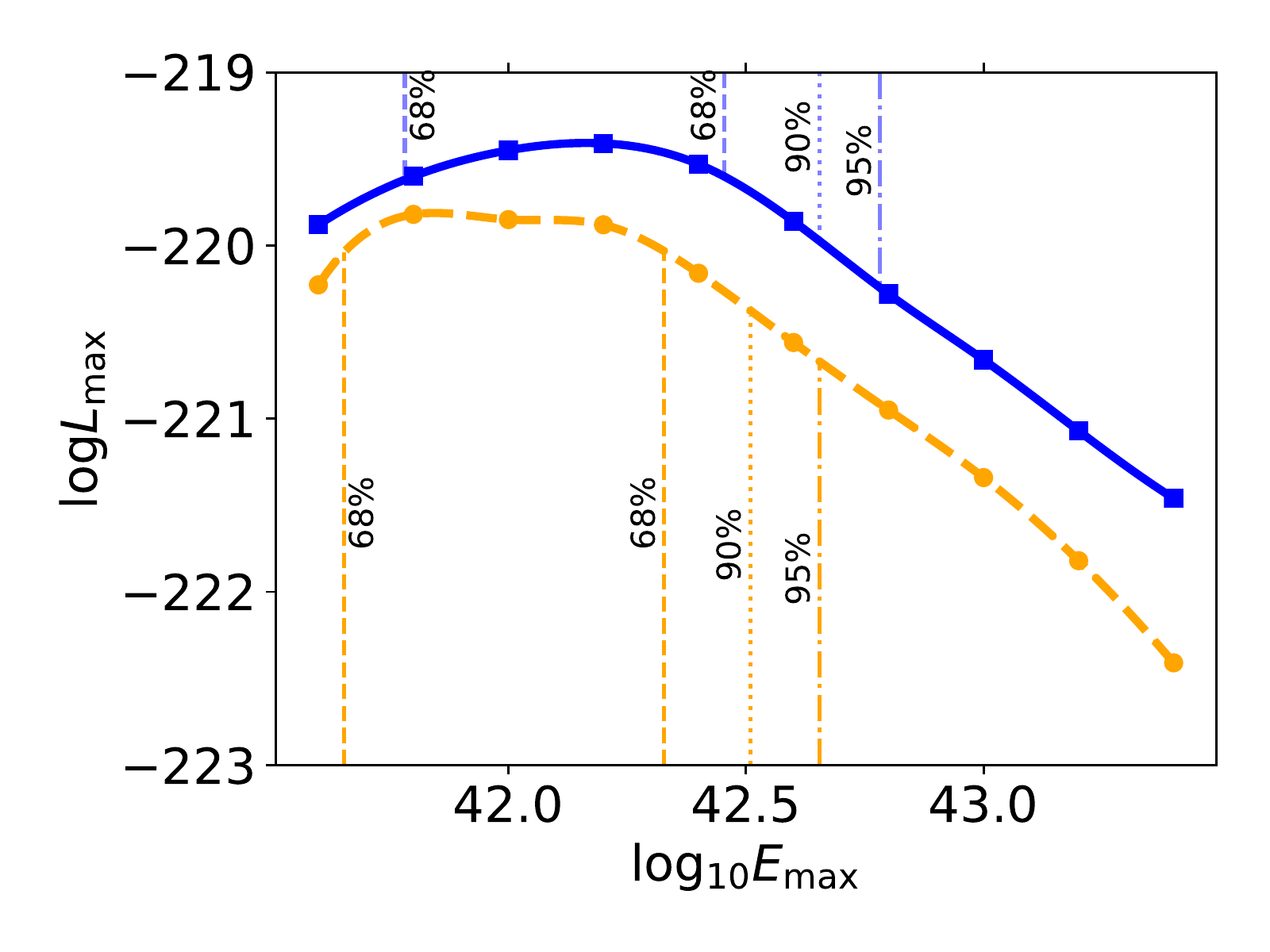}
     \includegraphics[width=0.32\textwidth]{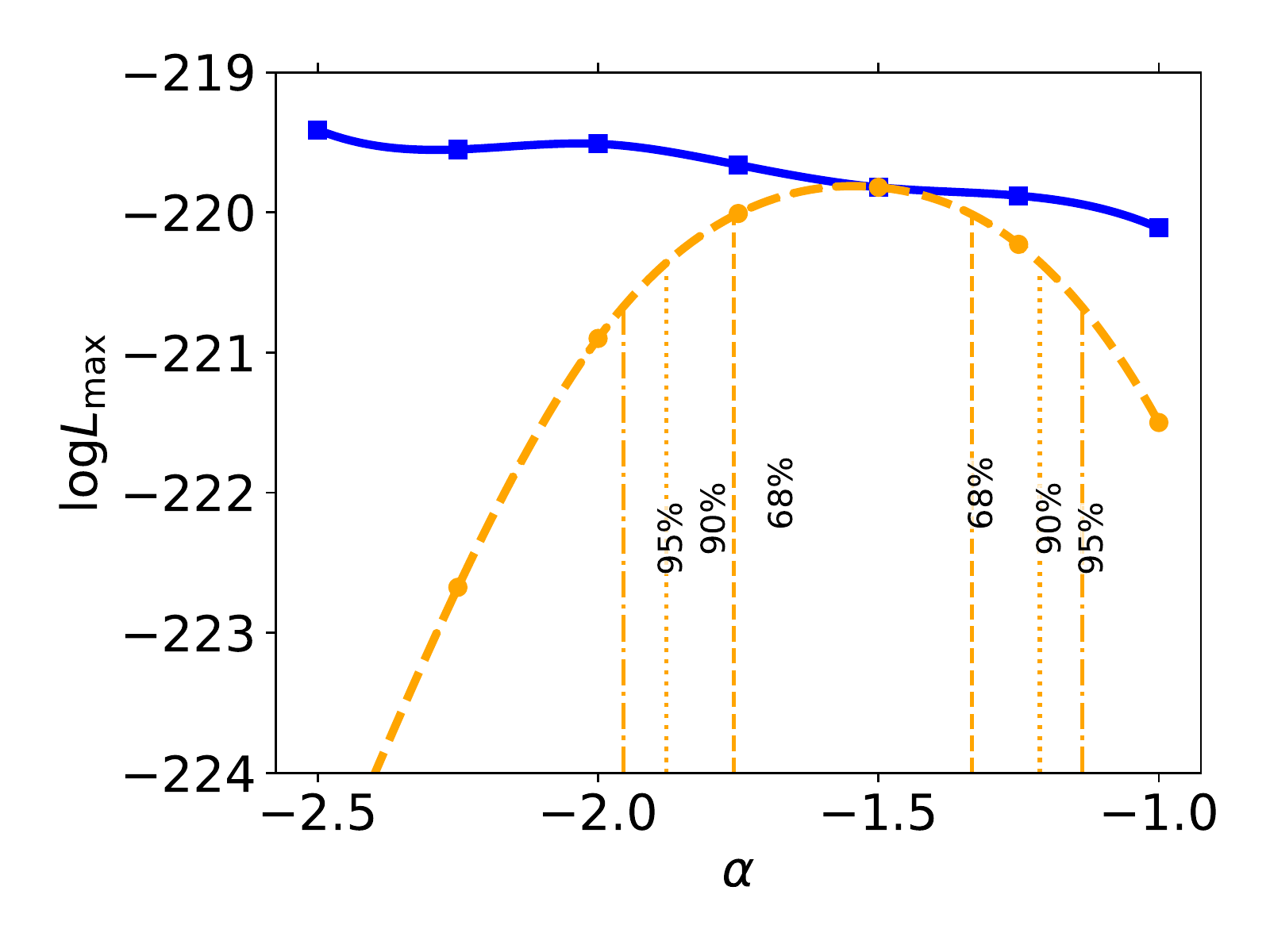}
      \includegraphics[width=0.32\textwidth]{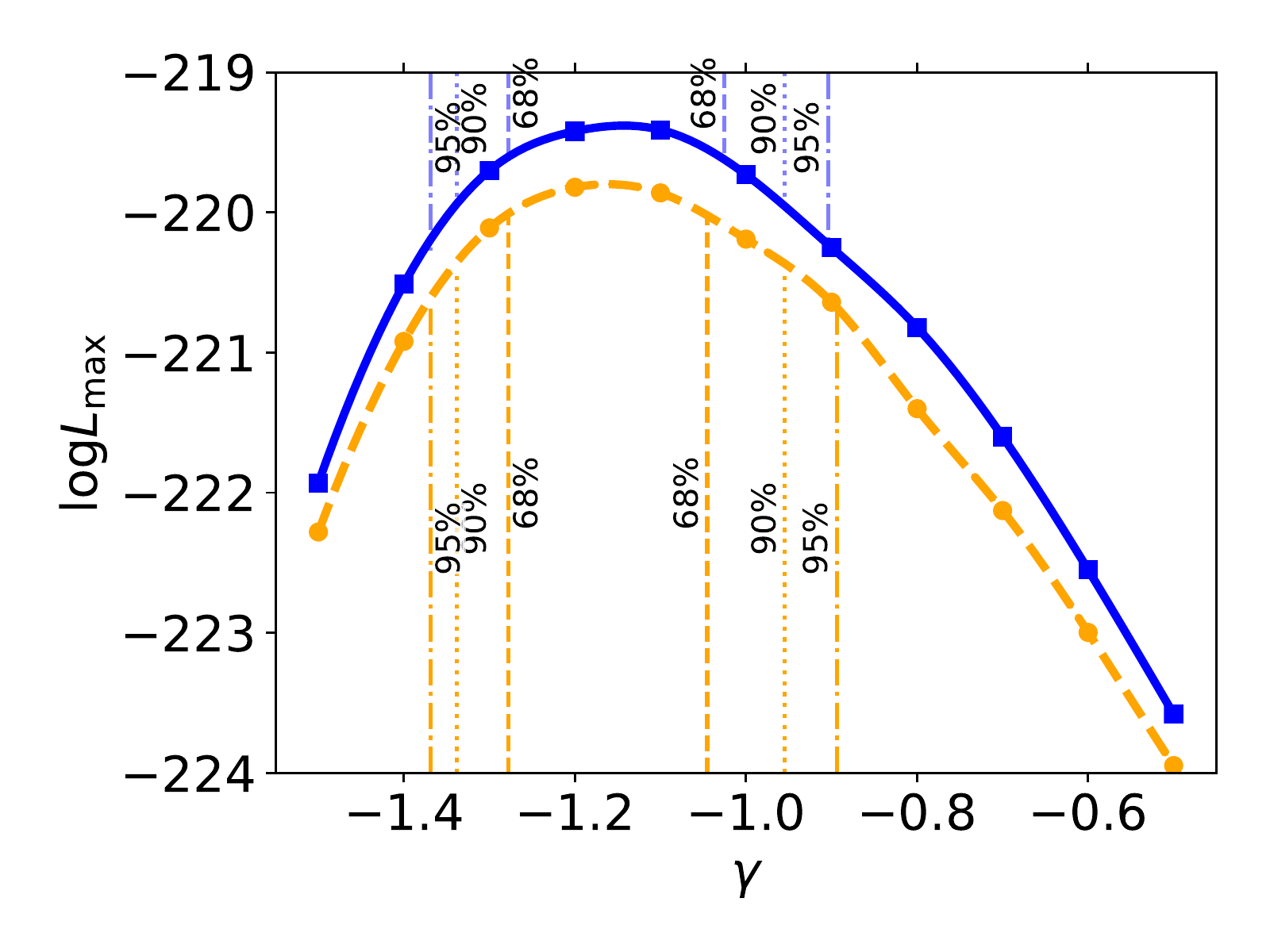} \\
       \includegraphics[width=0.32\textwidth]{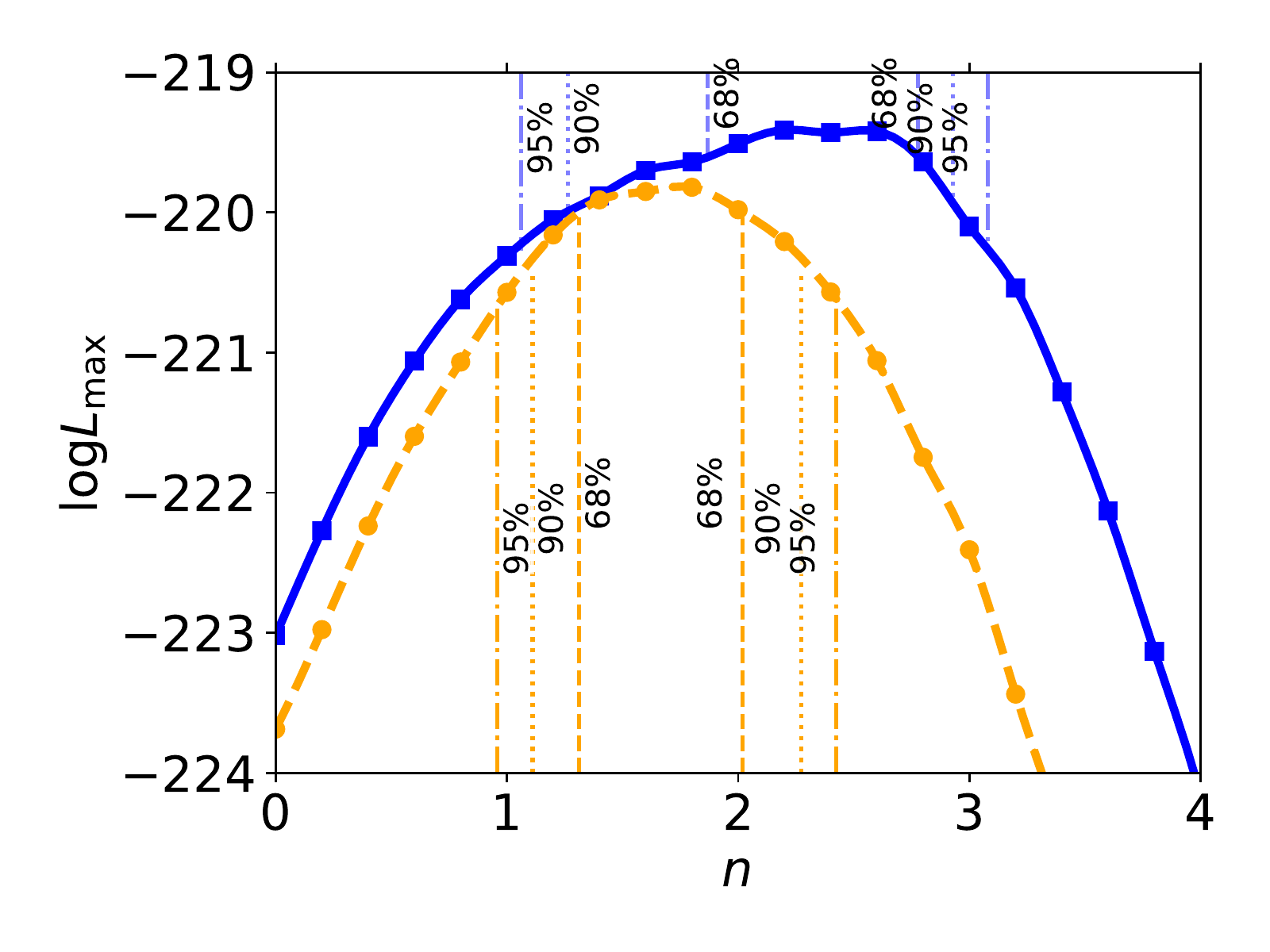}
        \includegraphics[width=0.32\textwidth]{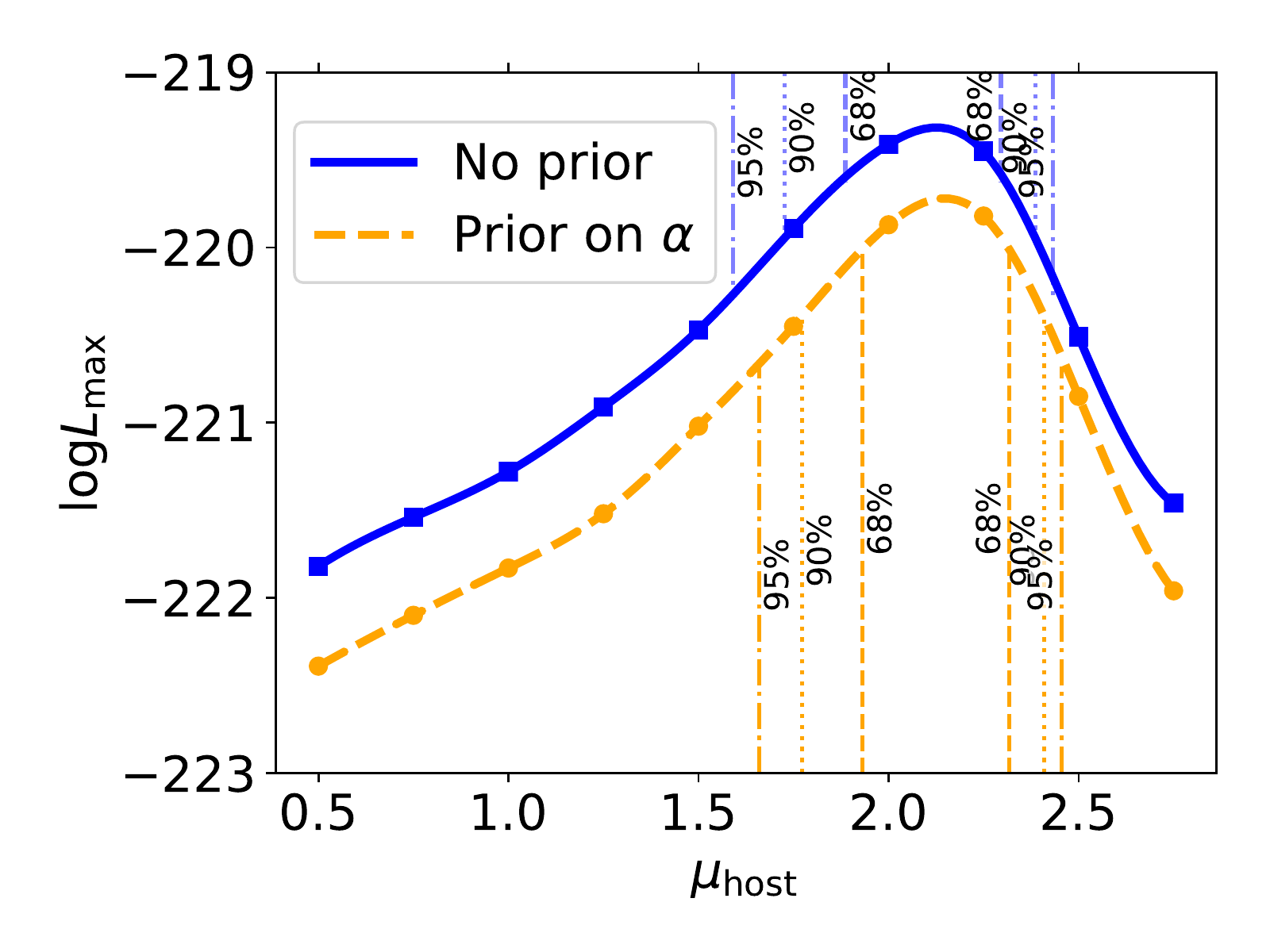}
         \includegraphics[width=0.32\textwidth]{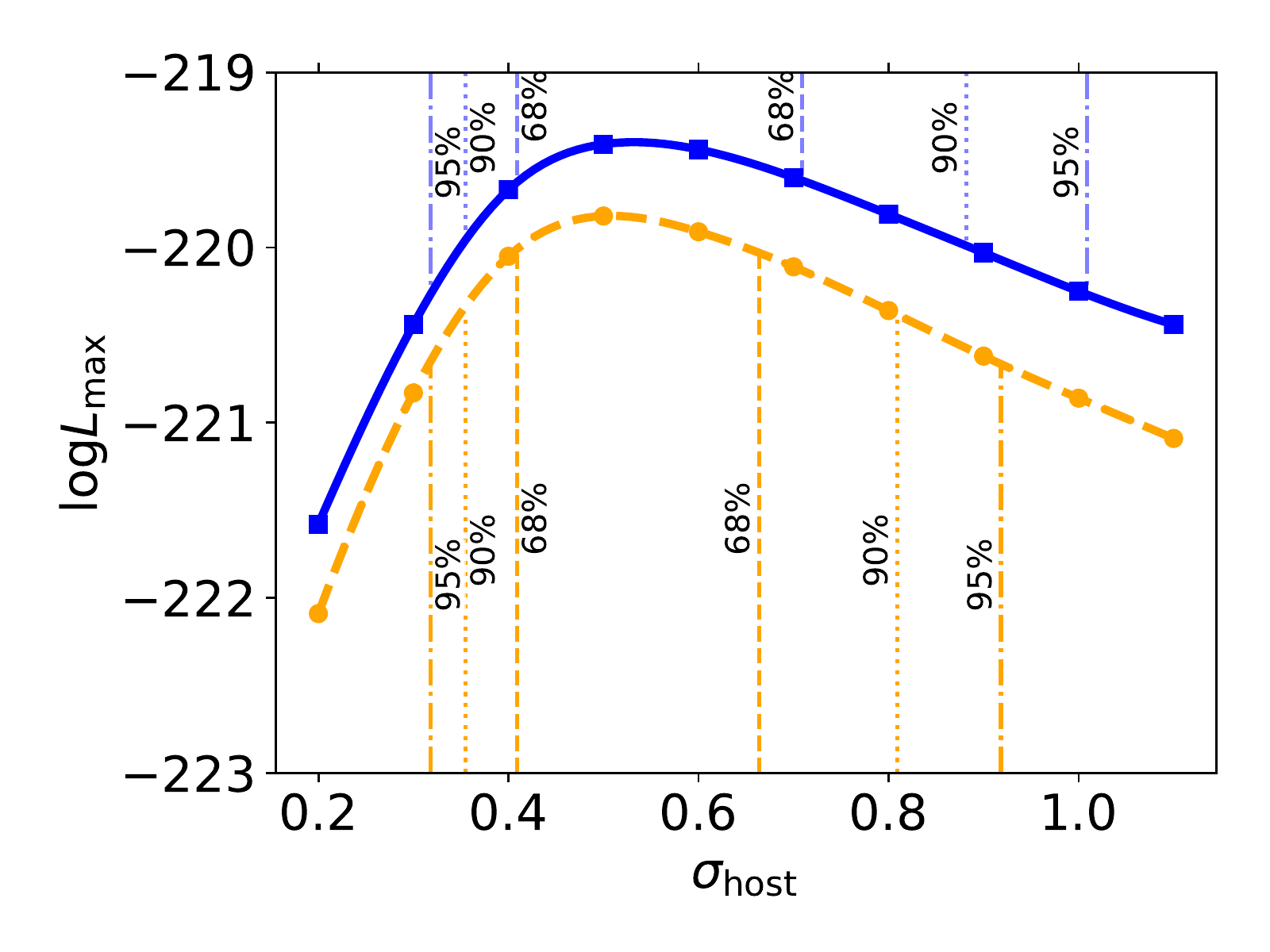}
    \caption{Maximum likelihoods as a function of each considered variable ($E_{\rm max}, \alpha, \gamma, n, \mu_{\rm host},\sigma_{\rm host}$) when marginalised over the other five, both with (orange, lower) and without (blue, upper) a prior on the spectral index $\alpha$. Calculation results are given by points, with lines drawn using cubic spline smoothing. Vertical lines are single-parameter intervals at the labelled degree of confidence calculated using Wilks' theorem with one degree of freedom. In the case of $\log_{10} E_{\rm max}$, 90\% and 95\% lower limits are at 41.4.}
    \label{fig:1dfigures}
\end{figure*}

Our single-parameter constraints are given in Figure~\ref{fig:1dfigures}, showing results both with and without a prior on $\alpha$. Best-fit values and confidence limits, calculated using Wilks' theorem \citep{Wilks62}, are given in  Table~\ref{tab:single_parameter_limits}. Two-parameter plots are given in Figure~\ref{fig:2dfigures}. We discuss the implications for each parameter individually below. 


\subsection{Maximum burst energy $E_{\rm max}$}
\label{sec:emax}

We find the maximum FRB to be energy is $\log_{10} E_{\rm max}$\,(erg)$= 41.84_{-0.18}^{+0.49}$ (68\% C.L.). According to our method, $E_{\rm max}$ is normalised to a bandwidth of 1\,GHz at the mean frequency of the data used as inputs (about 1350\,MHz), and applies to all burst widths. A strict lower limit on $E_{\rm max}$ is set by the intrinsically brightest localised FRB, 190711, which --- using a fluence of 34\,Jy\,ms \citep{Macquart2020}, 1\,GHz bandwidth, and $\alpha=-1.5$ --- had an energy of $E_{190711}=10^{41.5}$\,erg. $E_{\rm max}=10^{41.6}$\,erg is however consistent with observations at all quoted levels of confidence.

The preferred value of $E_{\rm max}$ is most strongly correlated with $\alpha$, which effectively attenuates FRBs as a function of redshift. Upper limits on $E_{\rm max}$ are also strongly correlated with $\gamma$, since a large negative value of this parameter makes it unlikely to observe FRBs near $E_{\rm max}$.

Our value of $E_{\rm max}$ lies in the middle of the values found by other authors. From Figure~\ref{fig:2dfigures}, fixing $n=0$ as per \citet{Luo2020} would lead to a lower value of $E_{\rm max}$, and greater consistency with that work. The higher values of $E_{\rm max}$ found by \citet{Lu2019}, and used by \citet{Arcus2020}, arise in models that assume a 1--1 DM--z relation, which will tend to over-estimate $E_{\rm max}$ when an FRB with a significant excess DM --- either due to its host or intervening matter --- is detected.

A key implication of $E_{\rm max}$ is the distance out to which an FRB is observable by a given telescope. For $\alpha=-1.5$, our value of $\log_{10} E_{\rm max}$ leads to a maximum observable redshift of $z=4^{+3}_{-0.85}$ for an instrument with 1\,Jy\,ms threshold.

\subsection{Intrinsic luminosity index $\gamma$}
\label{sec:gamma}

Our best-fit power-law index for the FRB population is $\gamma=-1.16_{-0.12}^{+0.11}$ (68\% C.L.).
As discussed by \citet{Macquart2018b}, this parameter primarily governs the degree to which FRBs are viewed from the near or far Universe, with steep values of $\gamma$ (i.e.\ below -1.5) leading to observations being dominated by nearby events and the event rate being governed by $E_{\rm min}$. Our result is definitely above this value, which is in agreement with all other calculations. It is however somewhat steeper than the values found by other authors.

Why? \citet{Luo2020} assume no cosmological source evolution, which this parameter is strongly correlated with. An increase of high-redshift FRBs can be due to either a lower $\gamma$, leading to more bursts visible near $E_{\rm max}$ in the larger volume of the distant Universe; or due to an evolving population, as determined by $n$. This anti-correlation is clearly visible in Figure~\ref{fig:2dfigures}. Both \citet{Arcus2020} and \citet{Lu2019} allow source evolution, but assume a 1--1 DM--z relation, implying a large distance for the highest-DM FRBs. In order to fit such bursts without over-predicting a large number of lower-energy bursts requires a flat luminosity function.

In the case that all FRBs repeat, with each FRB having the same $E_{\rm max}$ and $\gamma$ but a distribution of intrinsic rates, the intrinsic luminosity function for the entire population will match that of each FRB. This index has been well-measured for FRB~121102, with data giving a range $\gamma_{121102}\approx -0.9 \pm 0.2$ \citep{Law2017,Gajjaretal2018,James2019b}. This is consistent with our value for the population. However, should $E_{\rm max}$ vary over FRBs (which is quite likely), then the value of $\gamma$ for the population might be steeper. FRB observations by CHIME, which have detected several FRBs with many repeat bursts \citep{CHIME2019b,CHIME2019c,CHIME2020a}, should be able to answer the question definitively.

\subsection{Redshift evolution $n$}
\label{sec:n}

Our best-fit value of the redshift evolution scaling parameter is $n=1.77^{+0.25}_{-0.45}$ (68\% C.L.). Under the `rate interpretation' of $\alpha$, we find $n=1.26_{-0.35}^{+0.51}$. In both cases, $n=0$ is excluded at better than 99.9\%, which holds when no prior on $\alpha$ is considered.

Our detection of evolution in the FRB population supports conclusions based on FRB localisations, which locate most FRBs within normal host galaxies \citep{Heintz2020}; evidence associating FRBs with magnetars, such as the recent Galactic magnetar outburst \citep{Magnetar_CHIME,Magnetar_STARE2}; and observations of the host environment of FRB~121102 \citep{Michilli2018_121102}, as well as predictions from several classes of progenitor models \citep{Platts2018}.

This does not mean that we have confirmed that FRBs exhibit cosmological evolution identical to the star-formation rate however. A more-general model of source evolution, as used by \citet{Lu2019}, simply assumes a $(1+z)^{n^\prime}$ dependence, i.e.\ it removes the denominator and normalising constant in Eq.~\ref{eq:sfr_n}. Near $z=0$, $n^\prime \equiv 2.7 n$ --- however, the models will diverge above $z=1$. A true detection of scaling with the star-formation rate would require observations to be consistent with a downturn relative to the $(1+z)^{n^\prime}$ model at and beyond the peak of star-forming activity.

\begin{figure}
    \centering
    \includegraphics[width=\columnwidth]{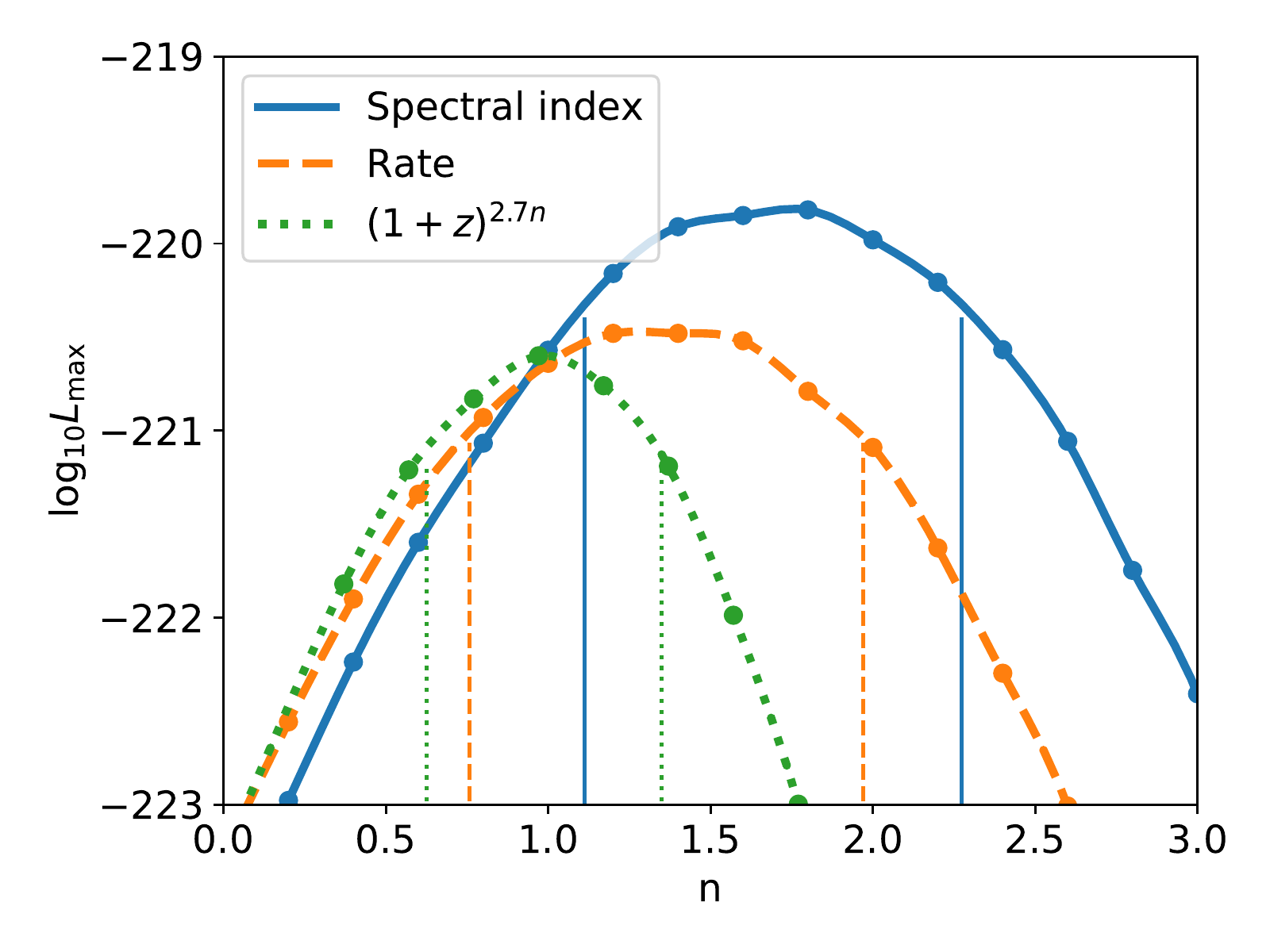}
    \caption{Maximum likelihood fits for source evolution parameter $n$, for three different cases: interpreting $\alpha$ as a spectral index, and source evolution $n$ scaling the star-formation rate as per Eq.~\ref{eq:sfr_n}
    (blue); interpreting $\alpha$ as a frequency dependent rate, again using Eq.~\ref{eq:sfr_n} for $n$-scaling (orange); and with $\alpha$ as a frequency-dependent rate, but $n$ scaling source evolution simply as $(1+z)^{2.7 n}$
    (green). The vertical lines show $90$ C.L.\ intervals calculated using Wilks'
     theorem.}
    \label{fig:compare_sfr}
\end{figure}

Figure~\ref{fig:compare_sfr} plots the likelihood for both interpretations of $\alpha$, and the $(1+z)^{2.7 n}$ model. While the spectral index interpretation of $\alpha$ clearly gives a better fit, the difference in maximum likelihoods between the two source evolution models under the rate assumption is negligible, with the preferred value of $n$ being slightly higher under SFR-scaling to compensate for the denominator in Eq.~\ref{eq:sfr_n}. We therefore cannot claim evidence for a down-turn in the source evolution function due to the peak in the SFR, only that the FRB population is evolving with cosmological time, with the rate per comoving volume greater at higher $z$.

Our result is still a significant improvement on prior works. Previous calculations have either had to assume a value for FRB source evolution (of $n=0$ or $1$), due to complete degeneracy with $\alpha$ \citep{Lu2019}, or otherwise could not distinguish between models \citep{Calebetal2016,Lu2019,Arcus2020}. As previously noted, and explained in detail in our companion paper, this degeneracy also affects this work. However, the degeneracy is not complete --- it is partially broken by the ASKAP/ICS sample of localised FRBs, and by the Parkes sample, which probes to sufficient $z$ to be sensitive to the non-Cartesian nature of the Universe. Of similar works, only \citet{Calebetal2016,Luo2020} model a telescope beamshape. We show in our companion paper that these authors' assumption of a Gaussian ($\sim$Airy) beamshape for Parkes observations is sufficient, but doing so for ASKAP data --- as considered by \citet{Luo2020} --- is inappropriate. The inability of \citet{Arcus2020} and \citet{Lu2019} to exclude $n=0$ may be due to their lack of beamshape modelling. Including beamshape reveals that a larger fraction of the sky is probed at lower sensitivity, thus increasing sensitivity to FRBs in the local Universe relative to that in the distant Universe. Without this effect, $n$ must be artificially decreased to model the observed number of near-Universe bursts.

\subsection{Excess DM distribution}
\label{sec:DMx}

Our model fits a log-normal distribution to DM$_{\rm host}$, which nominally covers the host galaxy and the immediate FRB environs. The fit will naturally include deviations from the NE2001 DM model of the Milky Way and the assumed halo DM of 50\,pc\,cm$^{-3}$. We find best-fit values of $\log_{10} \mu_{\rm host}=2.16^{+0.20}_{-0.23}$ and $\log_{10} \sigma_{\rm host}=0.51_{-0.10}^{+0.15}$, with both parameters being relatively independent of the other four. Figure~\ref{fig:2dfigures} shows that high values of $\mu_{\rm host}$ and low values of $\sigma_{\rm host}$ are most strongly excluded. The only other authors to fit these parameters are \citet{Macquart2020}, who use a sub-set of the data analysed in this work; our fitted value for the mean DM is greater than theirs, but not significantly. Partially, this is because \citet{Macquart2020} do not account for reduced sensitivity to high-DM bursts. Our inclusion of this effect requires a greater intrinsic high-DM population to fit the same observations. Combined with local contributions from the Milky Way's ISM and halo, 
we estimate a mean non-cosmological DM of ${\rm DM}_{\rm ISM} + {\rm DM}_{\rm halo} + \mu_{\rm host} = 50+35+145=230$\,pc\,cm$^{-3}$ at $z=0$. However, this still allows for low values of DM observed by ASKAP \citep{Shannonetal2018} and CHIME \citep{CHIME2019a}, since both DM$_{\rm host}$ and the cosmological contribution can vary. This large value of mean non-cosmological contribution helps to explain the observation by \citet{Shannonetal2018} that the mean DM of the Parkes FRB sample is not as large relative to the ASKAP/ICS sample as would be expected from the relative telescope sensitivities alone.

\subsection{The prevalence of FRBs}

We estimate the best-fit absolute rate of FRBs above $E_{\rm min}$, $\Phi_0$, by maximising the product of $p_n$ between the ASKAP/FE and Parkes/Mb samples with well-constrained $T_{\rm obs}$.
We quote $\Phi_{39}$, defined as the estimated rate of bursts above $10^{39}$\,erg \citep[above the maximum allowed value of $E_{\rm min}$][]{James2021Full} per year at $z=0$. In the case of our best-fit model, we find $\Phi_{39}=9_{-3.8}^{+2.2}\cdot 10^4$\,bursts Gpc$^{-3}$\,yr$^{-1}$ (90\% C.L.).

This value is broadly consistent with that estimated by other authors \citep{RaviNature2019,Lu2019,Luo2020}, and supports the conclusion that the majority of FRBs must either be repeaters, or cannot be due to known populations of once-off events.

Interestingly, the best-fit parameter set under-predicts the number of FRBs observed by ASKAP/FE (12.9 vs.\ 20 in 1274.6 days), and over-predicts the number found by Parkes (17.0 vs 12 in 164.4 days). There are several possible causes of this discrepancy. One possibility is a minimum FRB energy --- or at least a flattening of the distribution at low energies --- which would reduce the number of bursts seen by the more-sensitive Parkes telescope. Another is the low number of FRBs detected by Parkes with SNR below 16, as noted by \citet{Jamesetal2019a}, which could be an indicator of a reduced detection efficiency to low-fluence bursts. Both are investigated and deemed unlikely in our companion paper. A third option is that the observation times reported here are raw observation times, and do not account for lost effective observation time due to e.g.\ radio-frequency interference, which is likely more prevalent at Parkes than ASKAP. Finally, this could be the result of simple statistical fluctuations --- if the true rates are 12.9 and 17.0 FRBs respectively, then the product of the ASKAP and Parkes likelihoods will be this or less unlikely 7.7\% of the time.

\section{Conclusion}
\label{sec:conclusion}

We have used a precise and accurate method of modelling the results of FRB surveys to fit the measured DM, $z$, and signal-to-noise ratios of FRBs detected by ASKAP and Parkes. We have carefully selected our data to ensure it is not biased due to under-reporting of observation time, or due to large local DM contributions reducing sensitivity. Crucially, we have included a sample of localized FRBs from ASKAP for which the redshift of the host galaxies is measured.

These modelled observations are tested against a six-parameter model of the FRB population. Using a maximum-likelihood approach, we have derived the tightest constraints on FRB population parameters to date. Our value of the maximum FRB energy of $41.84_{-0.18}^{+0.49}$\,erg (68\% C.L.) is mid-way between previous estimates.
The intrinsic slope of the cumulative luminosity distribution, $\gamma$, is found to be $-1.16_{-0.12}^{+0.11}$ (68\% C.L.), consistent with, but slightly steeper than, the slope found for FRB~121102. Importantly, we find that the FRB population evolves with redshift, scaling with the star-formation rate (SFR) to the power of $1.77_{-0.45}^{+0.25}$ or $1.26_{-0.35}^{+0.51}$, depending on the interpretation of FRB spectral properties. While we cannot distinguish between SFR-scaling and a model where the FRB population increases as a simple power of $(1+z)^{2.7 n}$, in all scenarios we exclude a non-evolving population at better than 99.9\% C.L. 

Our best-fit log-mean host contribution to DM of $145$\,pc\,cm$^{-3}$ is also somewhat higher than the standard value of $100$\,pc\,cm$^{-3}$.

Such large excess dispersion measures, and a population evolution consistent with star formation, strongly aligns with the hypothesis of FRBs originating from young magnetars. We caution that these results apply to the total FRB population (which may or may not consist of multiple sub-populations), and only to that part of the population to which the ASKAP and Parkes observations are sensitive.

\section*{Acknowledgements}

This research has made use of NASA's Astrophysics Data System Bibliographic Services. This research made use of Python libraries \textsc{Matplotlib} \citep{Matplotlib2007}, \textsc{NumPy} \citep{Numpy2011}, and \textsc{SciPy} \citep{SciPy2019}. This work was performed on the gSTAR national facility at Swinburne University of Technology. gSTAR is funded by Swinburne and the Australian Government’s Education Investment Fund. This work was supported by resources provided by the Pawsey Supercomputing Centre with funding from the Australian Government and the Government of Western Australia. This research was partially supported by the Australian Government through the Australian Research Council's Discovery Projects funding scheme  (project DP180100857).

\section*{Data Availability}

The data underlying this article will be shared on reasonable request to the corresponding author.


\bibliographystyle{mnras}
\bibliography{bibliography}

\appendix

\newpage
\section{Further data}
\label{sec:extra_data}

\begin{figure*} 
    \centering
    \includegraphics[width=0.32\textwidth]{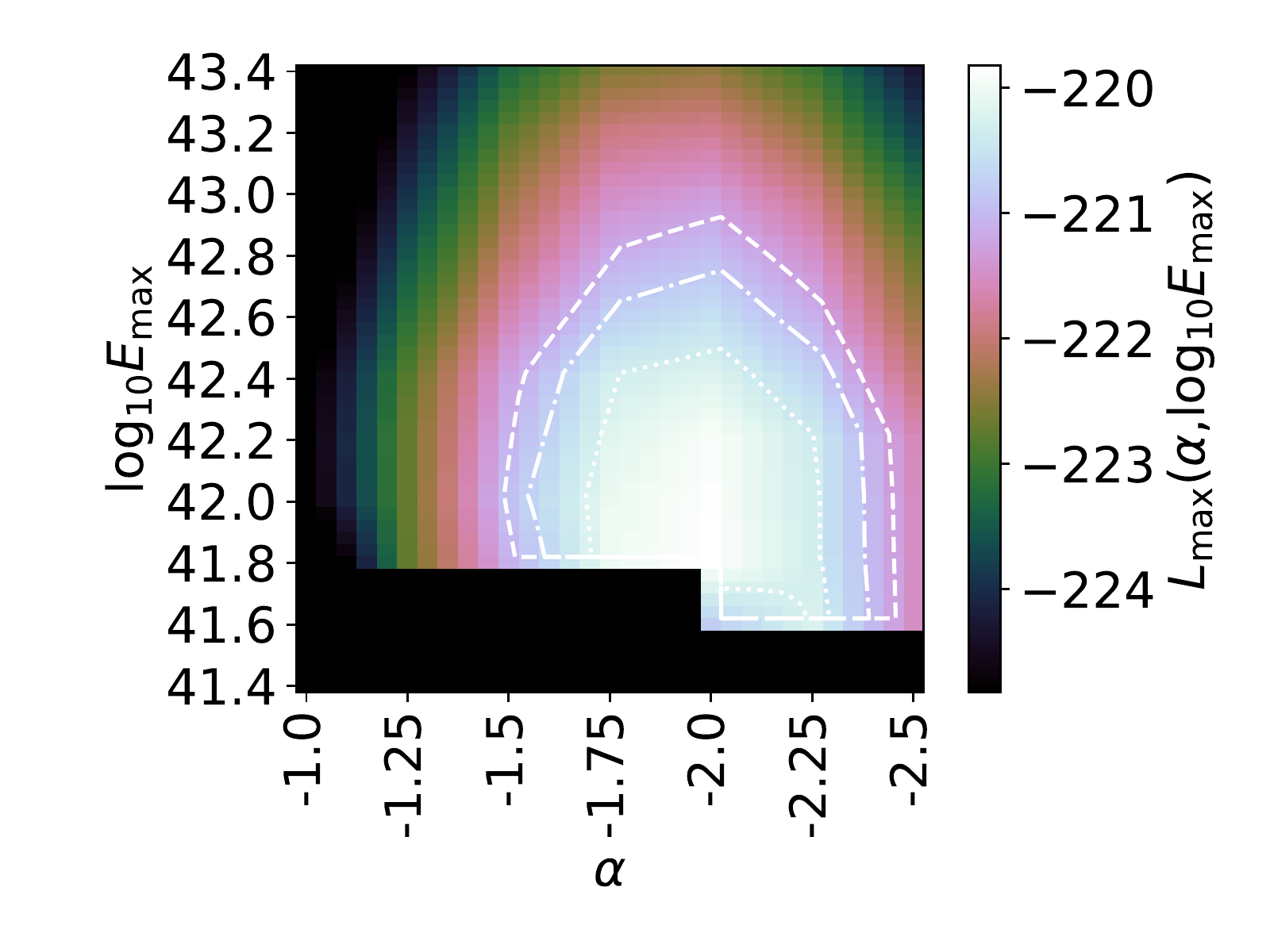}
     \includegraphics[width=0.32\textwidth]{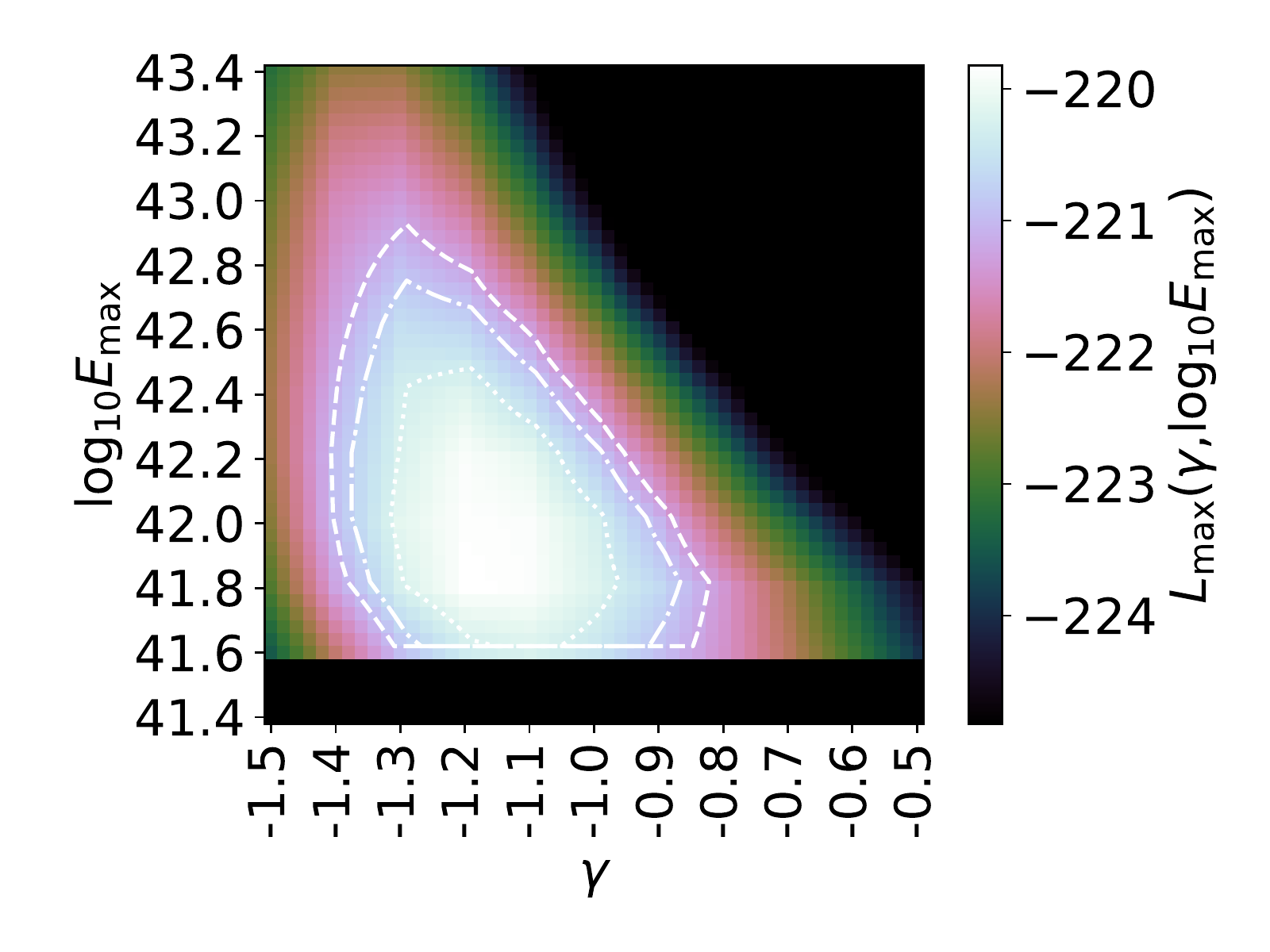}
      \includegraphics[width=0.32\textwidth]{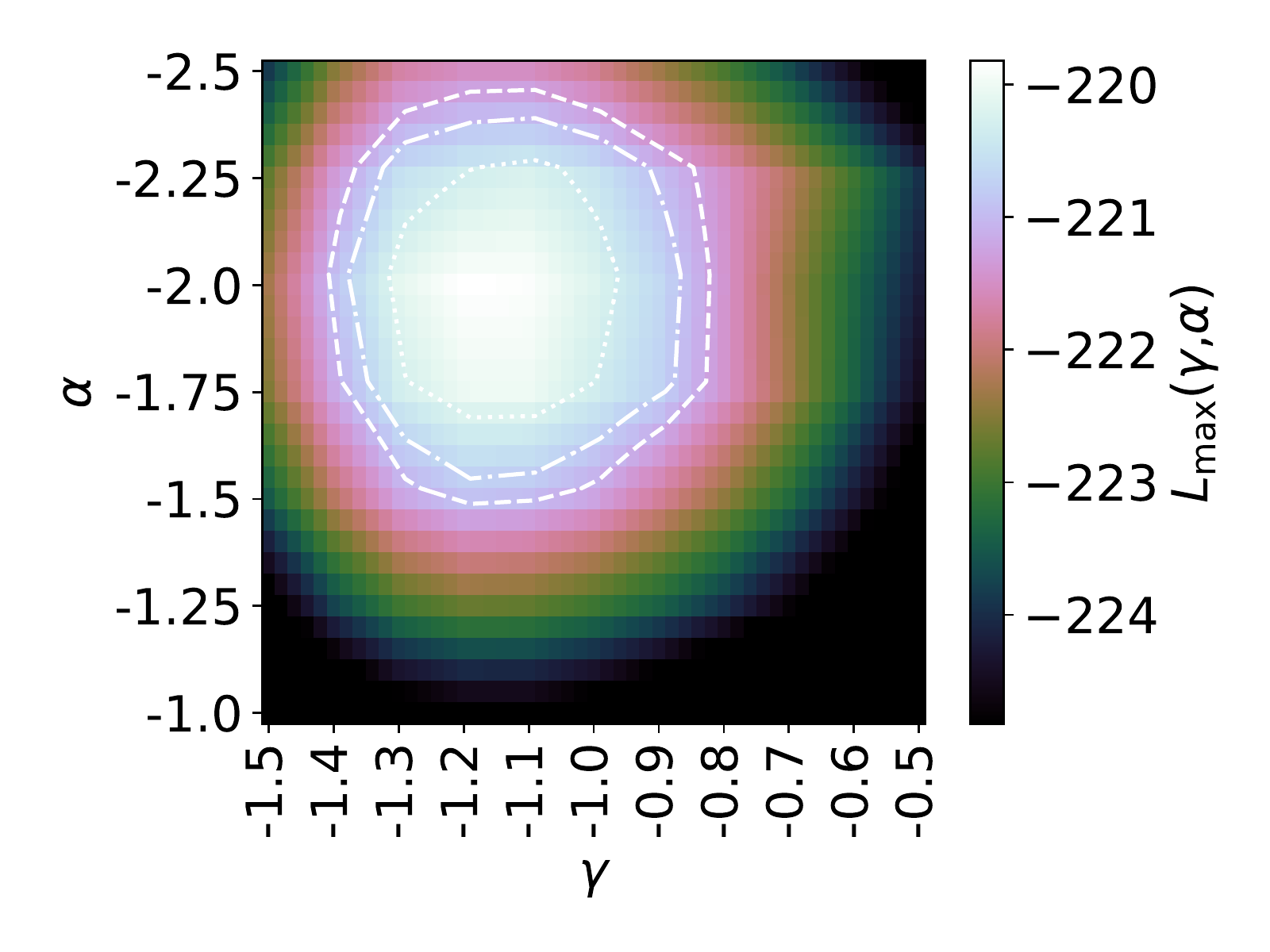} \\
       \includegraphics[width=0.32\textwidth]{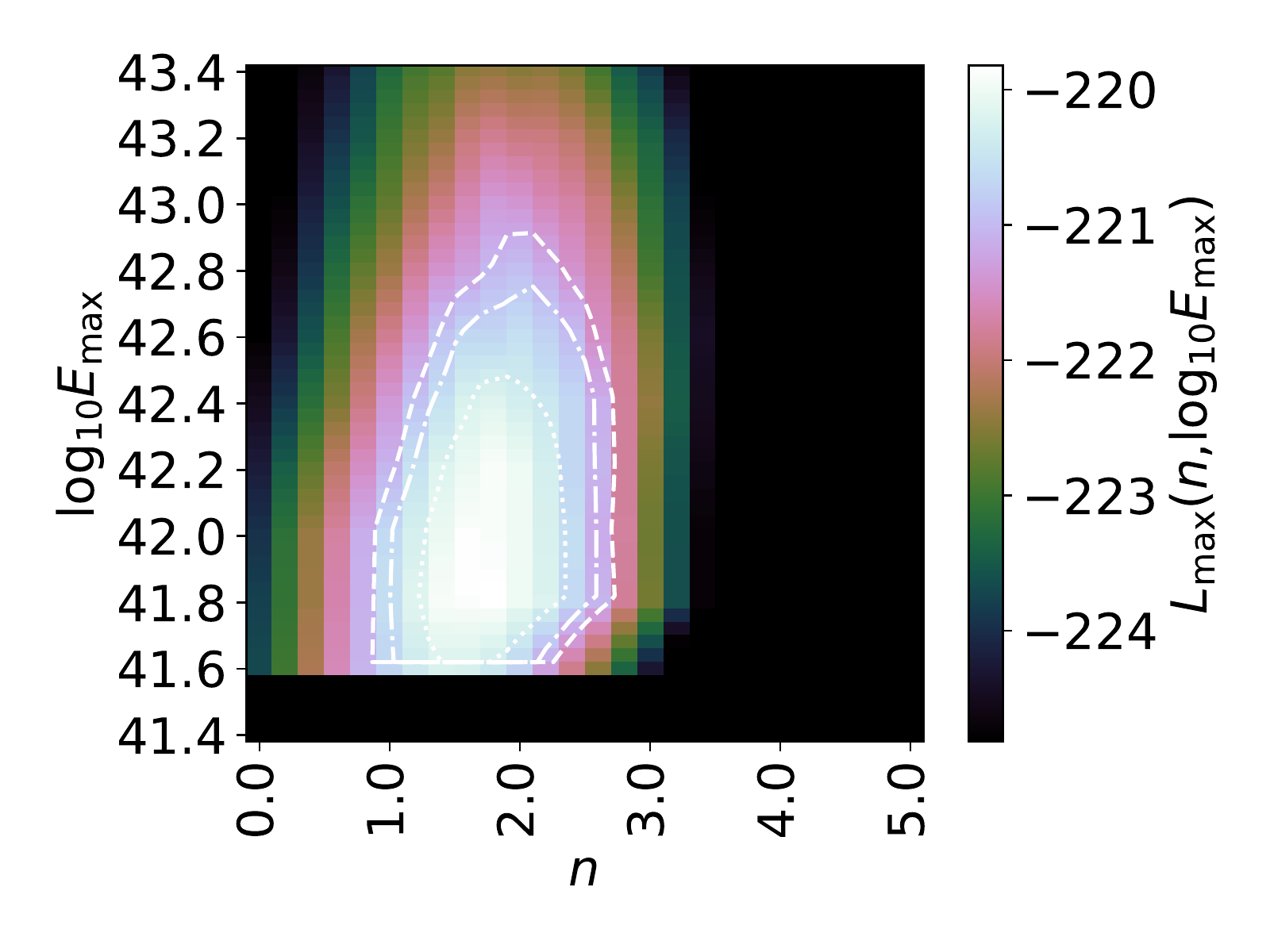}
        \includegraphics[width=0.32\textwidth]{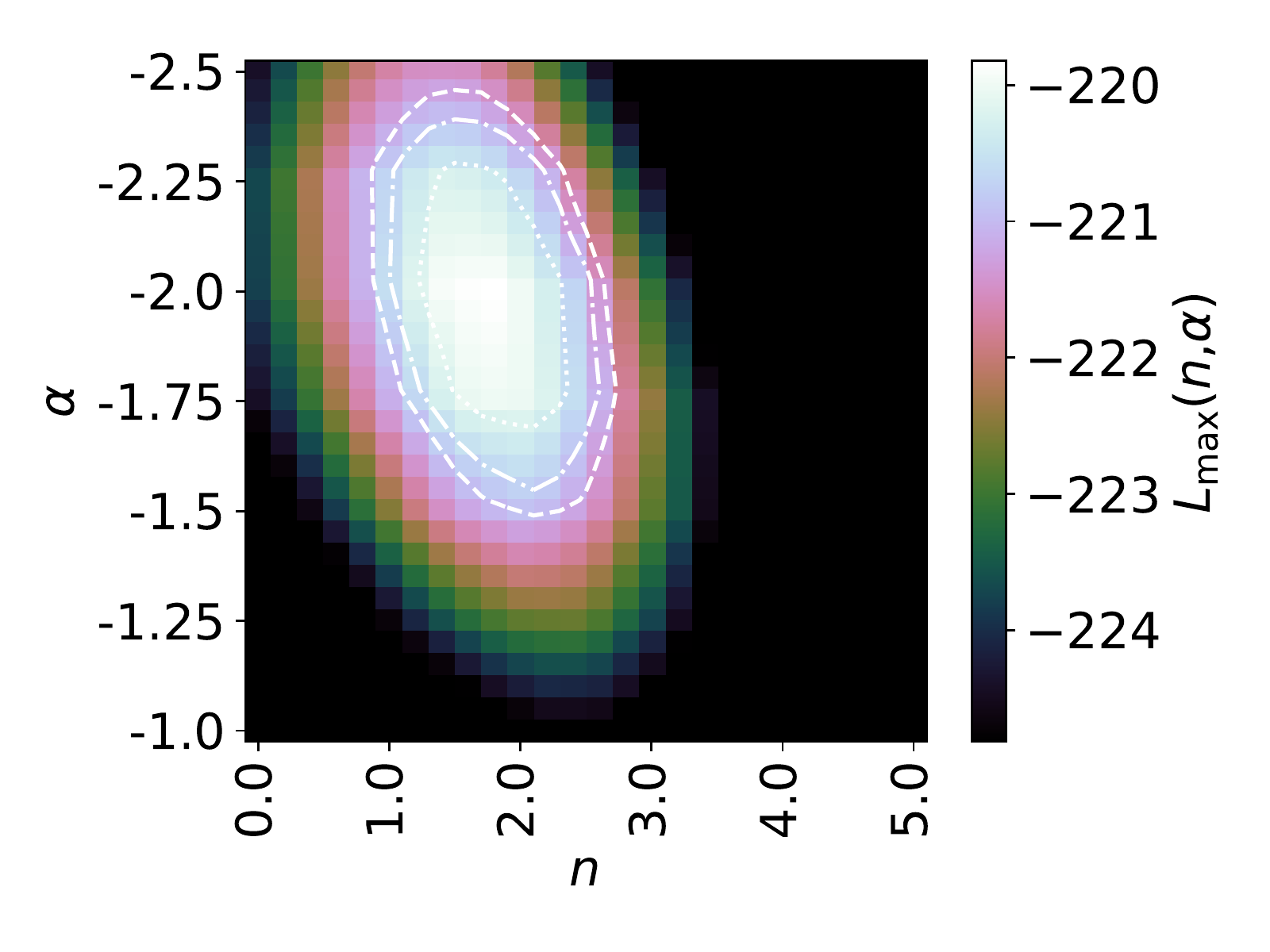}
       \includegraphics[width=0.32\textwidth]{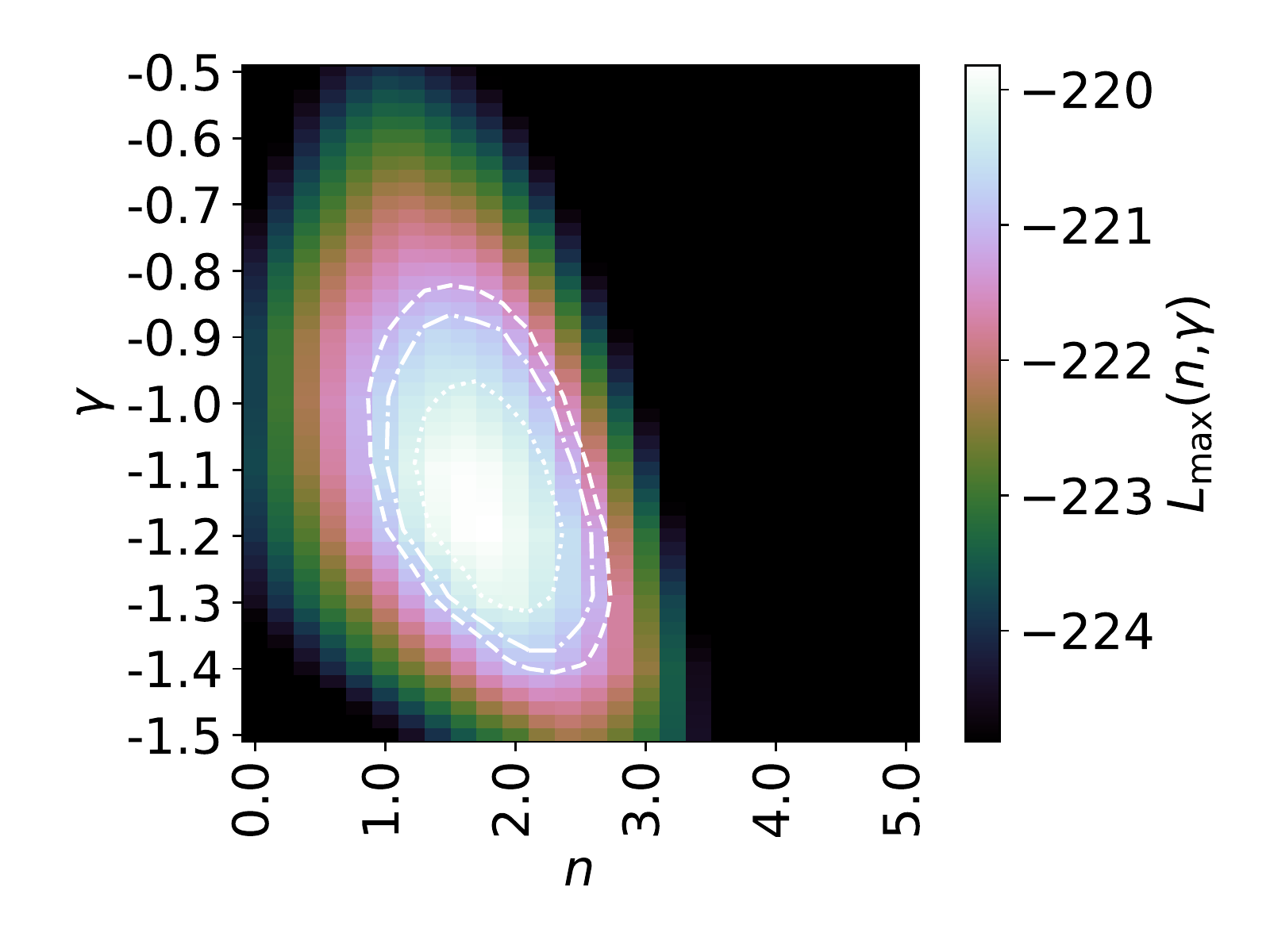} \\
        \includegraphics[width=0.32\textwidth]{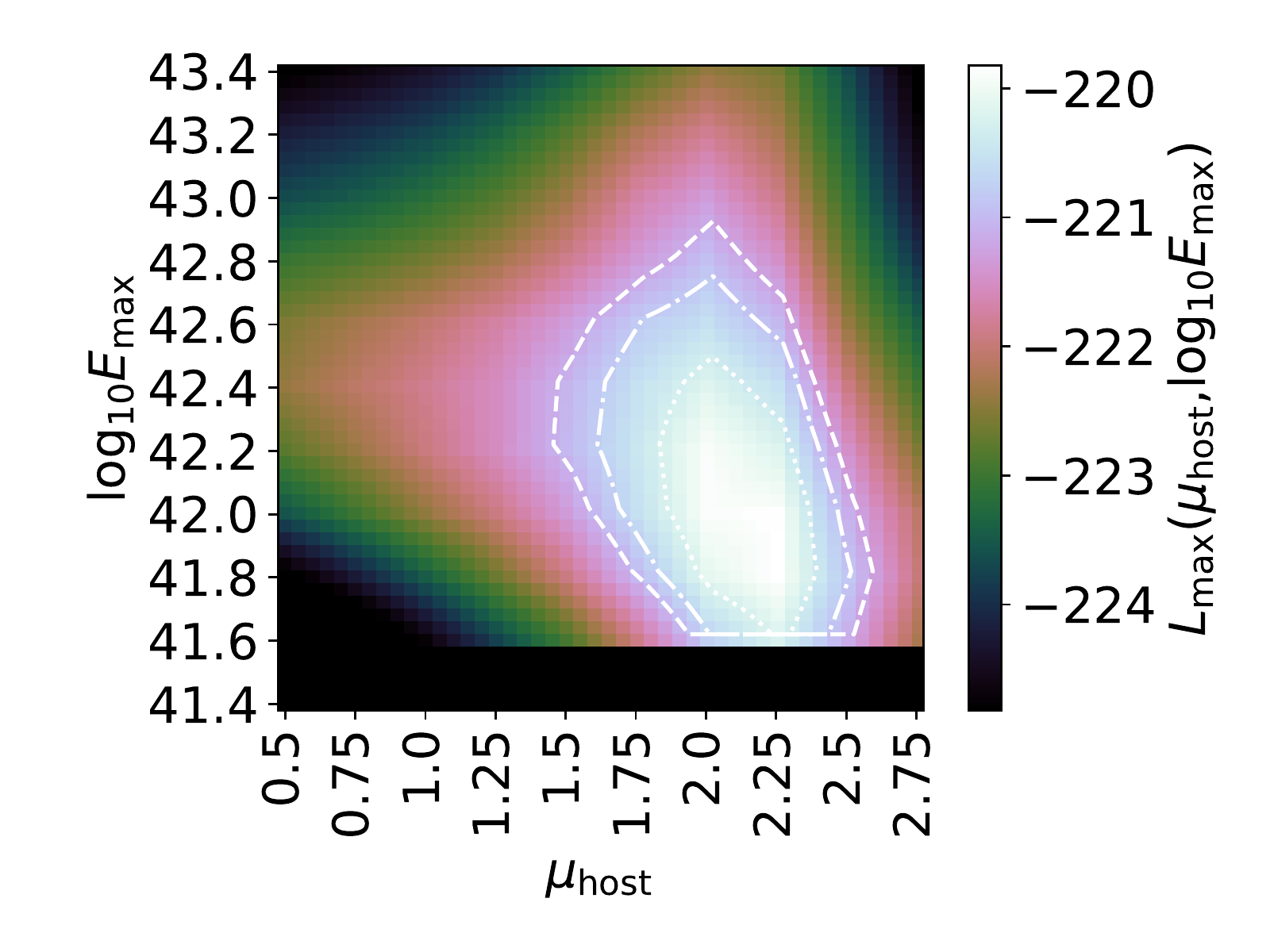}
       \includegraphics[width=0.32\textwidth]{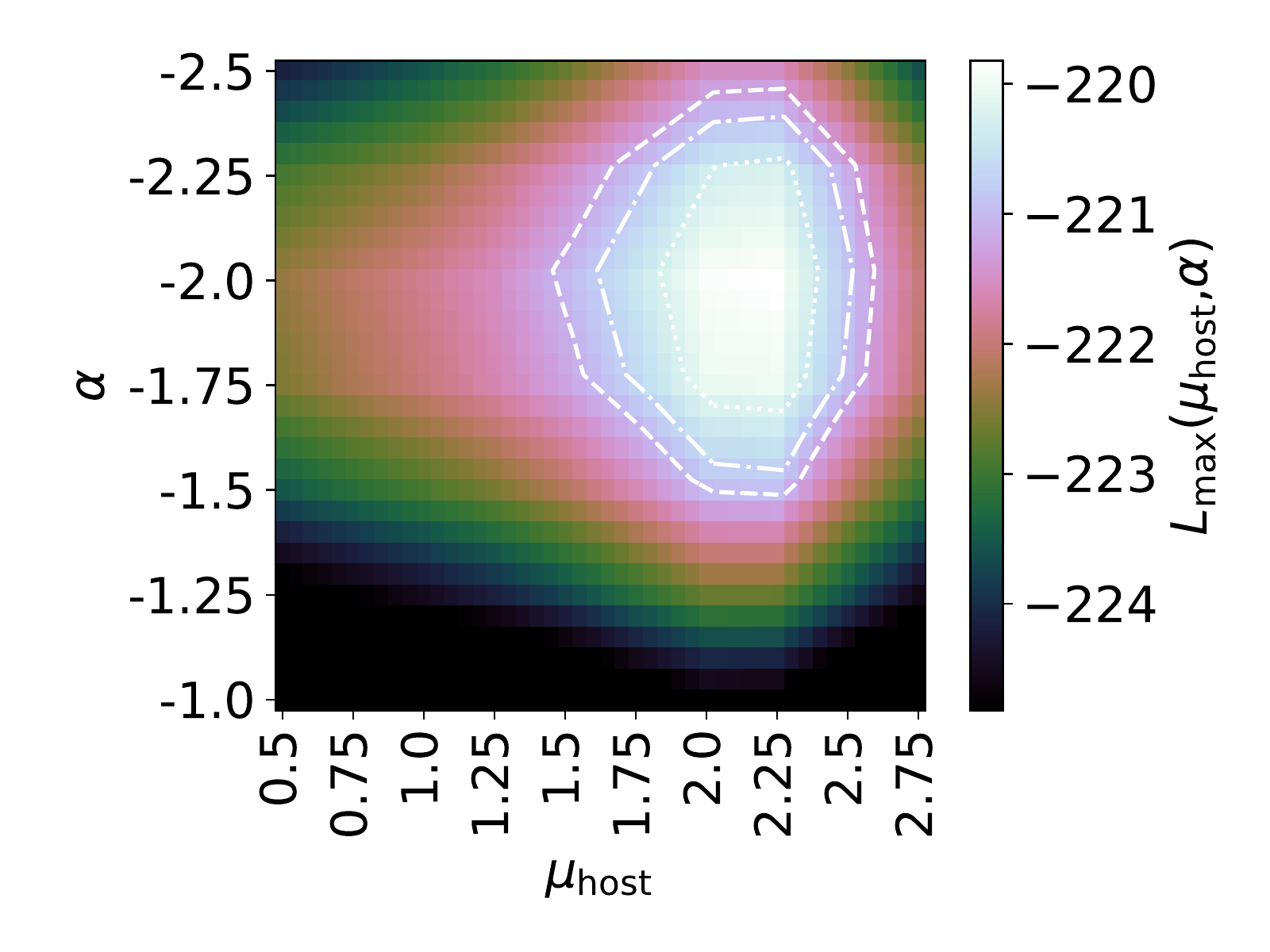}
        \includegraphics[width=0.32\textwidth]{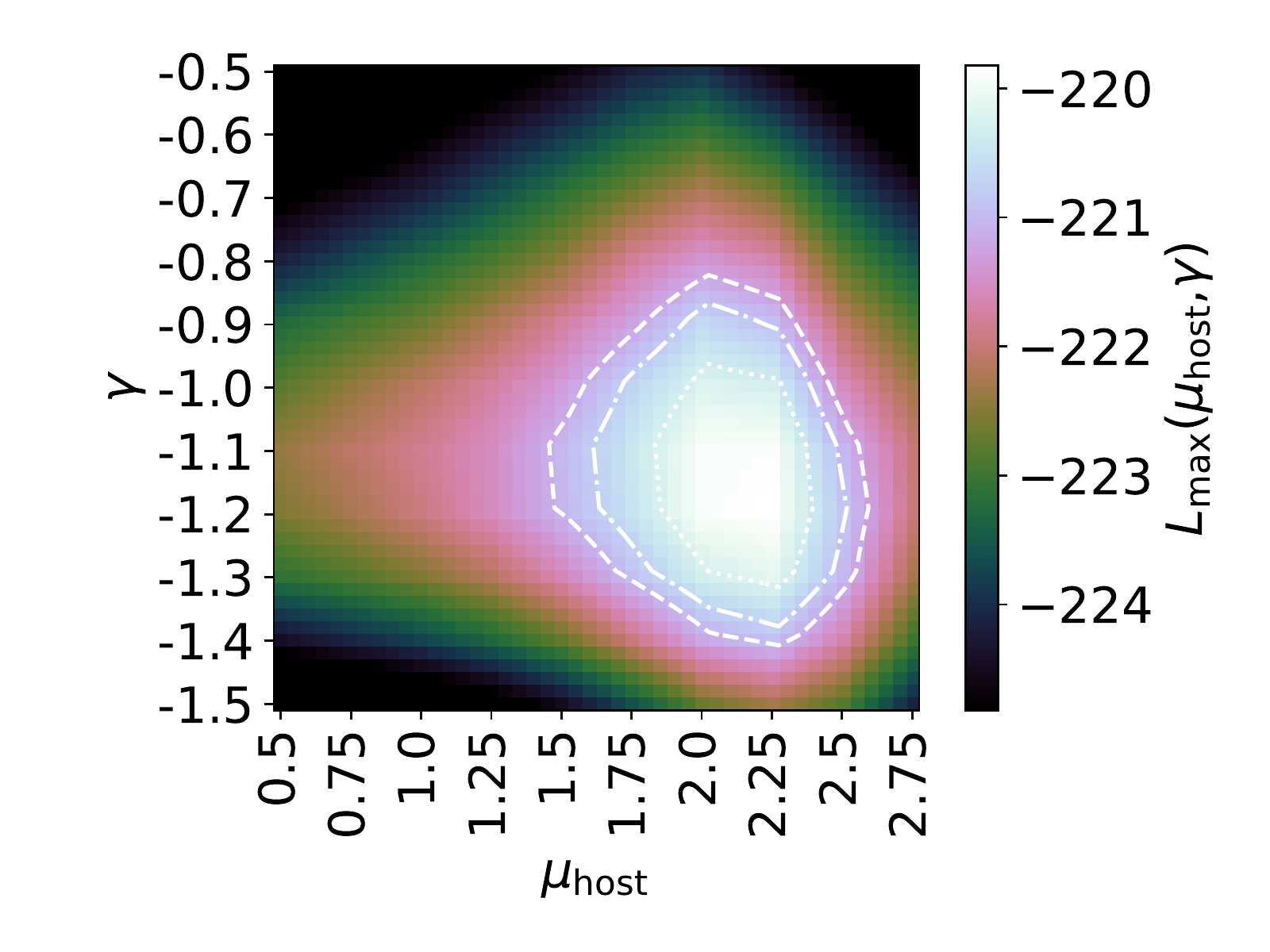}\\
        \includegraphics[width=0.32\textwidth]{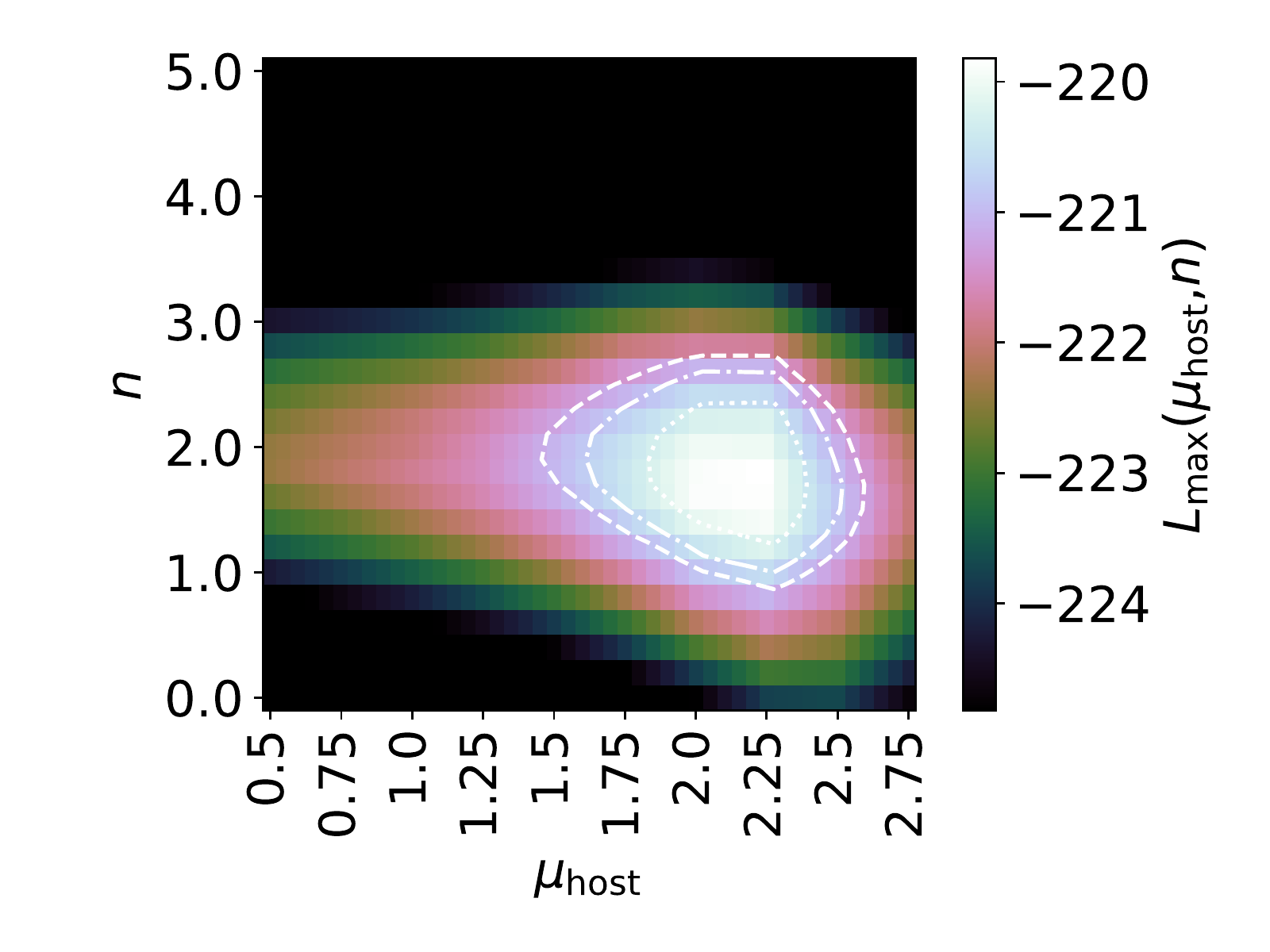}
        \includegraphics[width=0.32\textwidth]{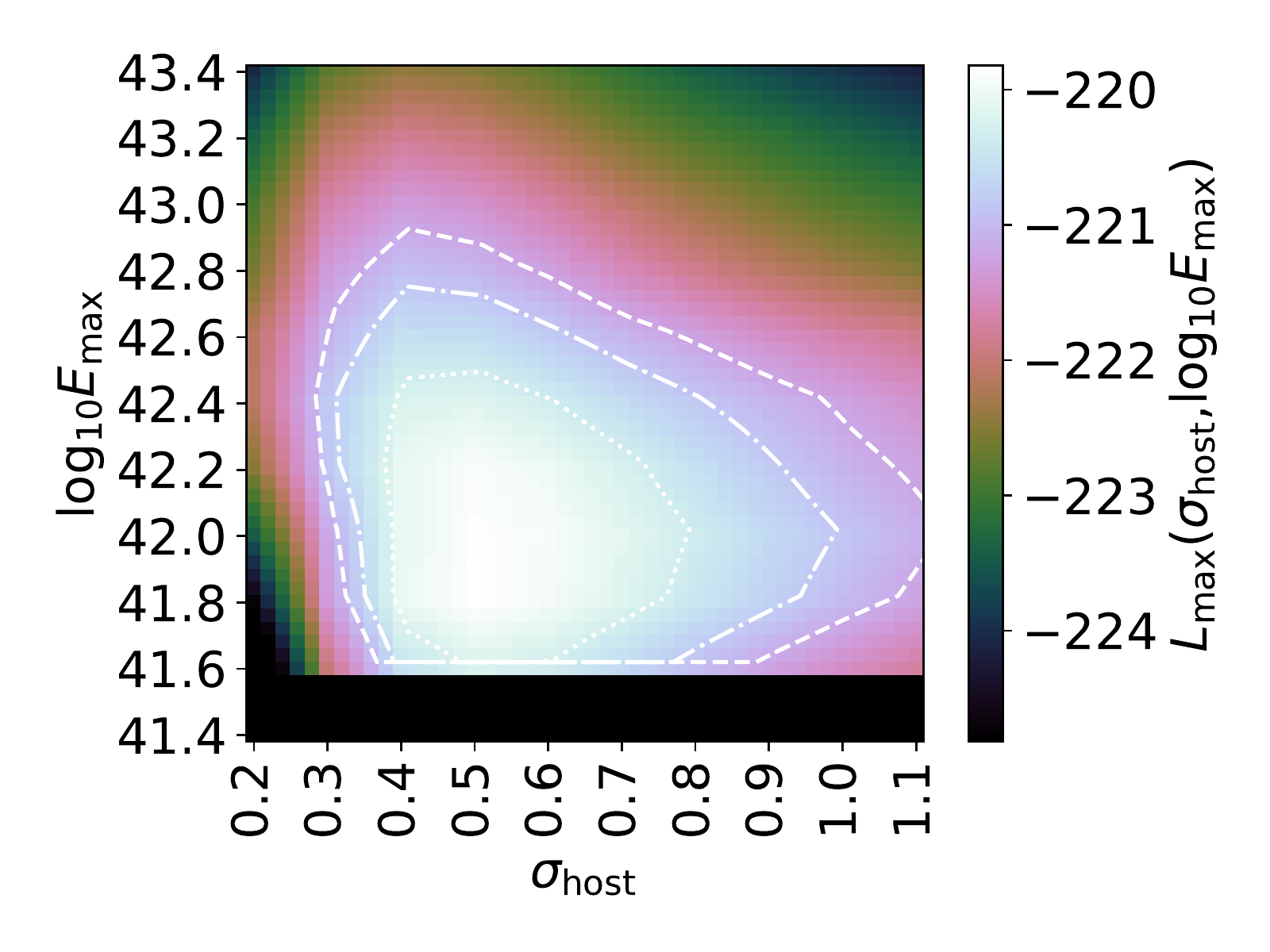}
       \includegraphics[width=0.32\textwidth]{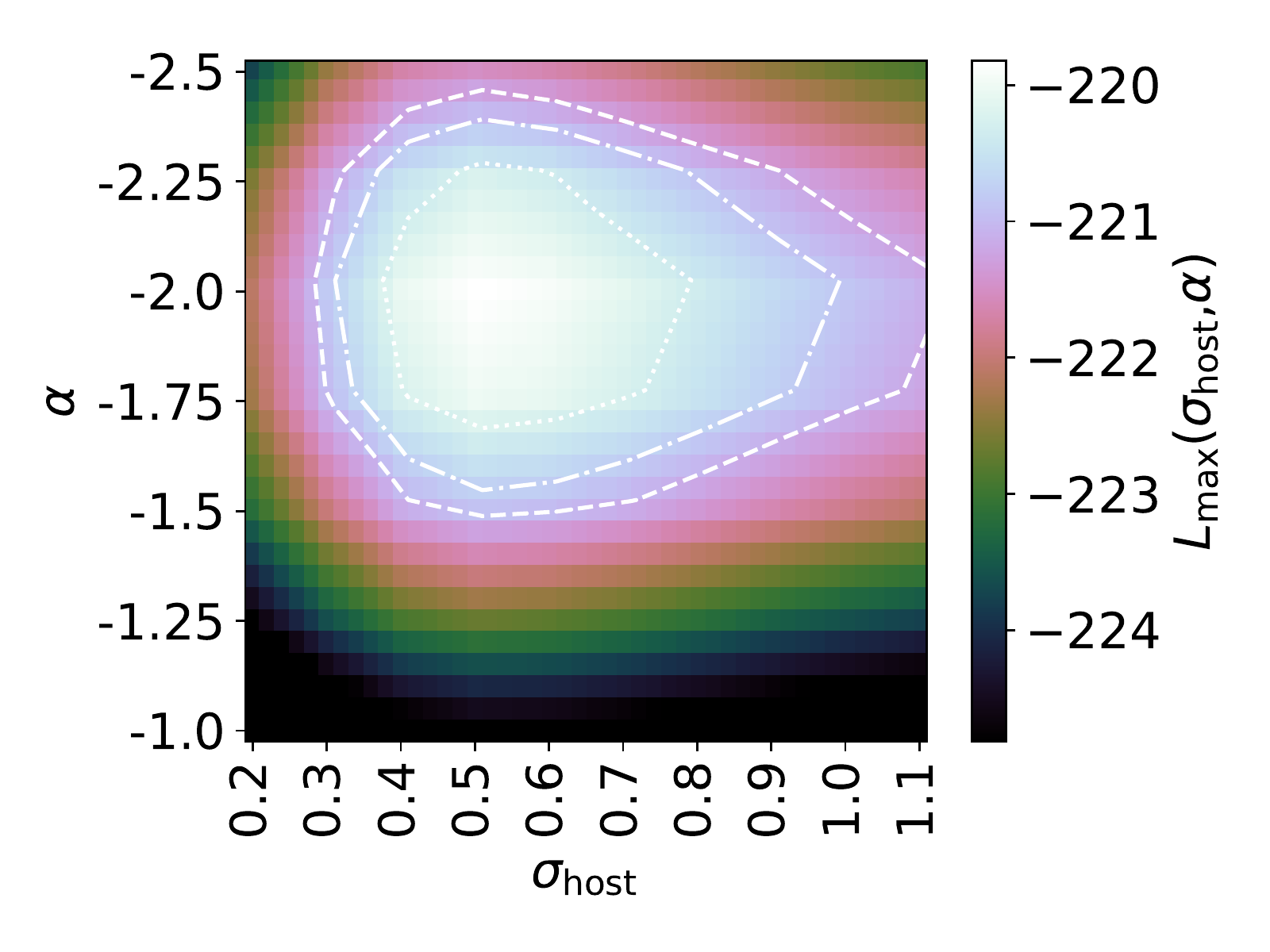} \\
        \includegraphics[width=0.32\textwidth]{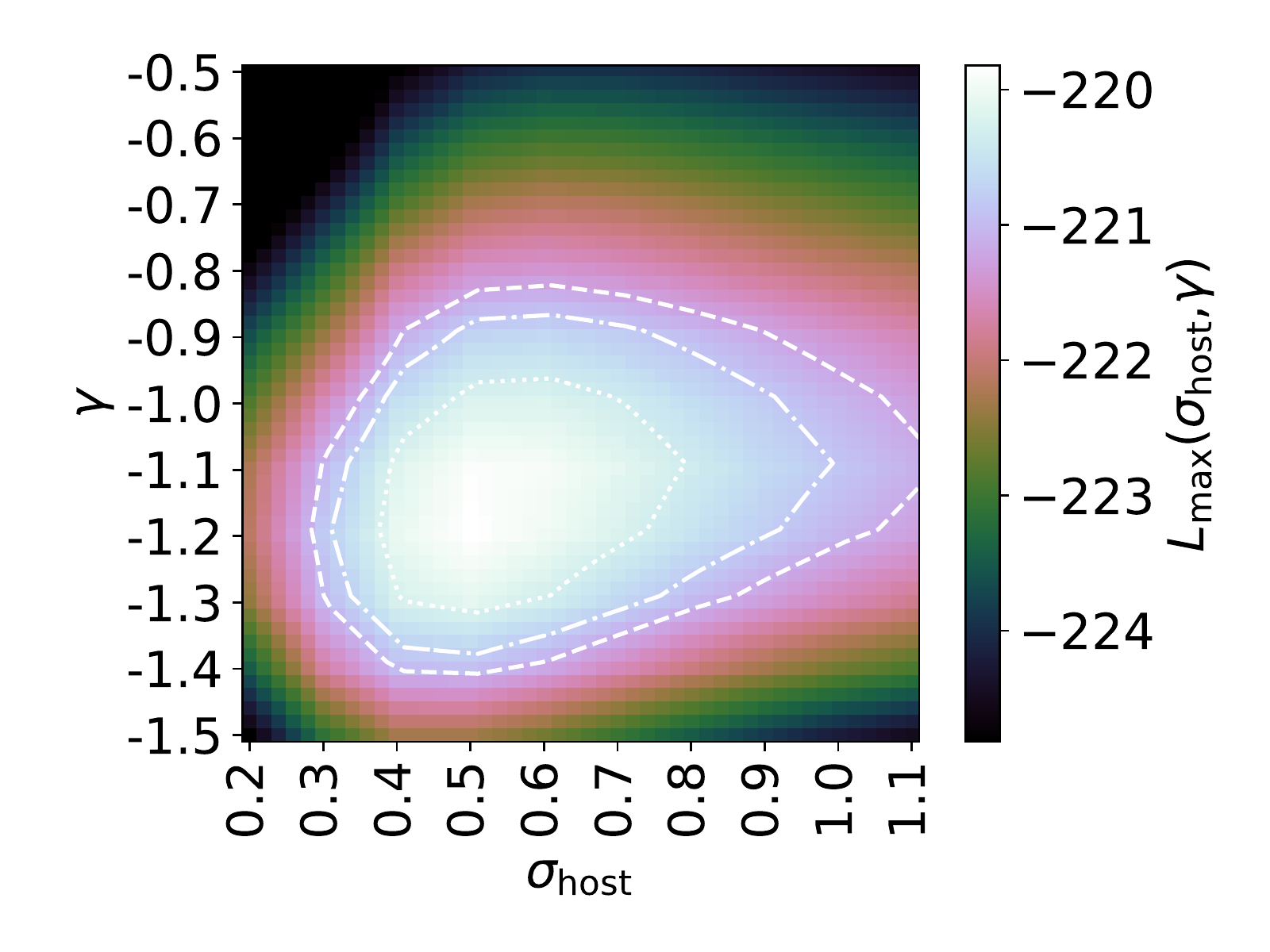}
       \includegraphics[width=0.32\textwidth]{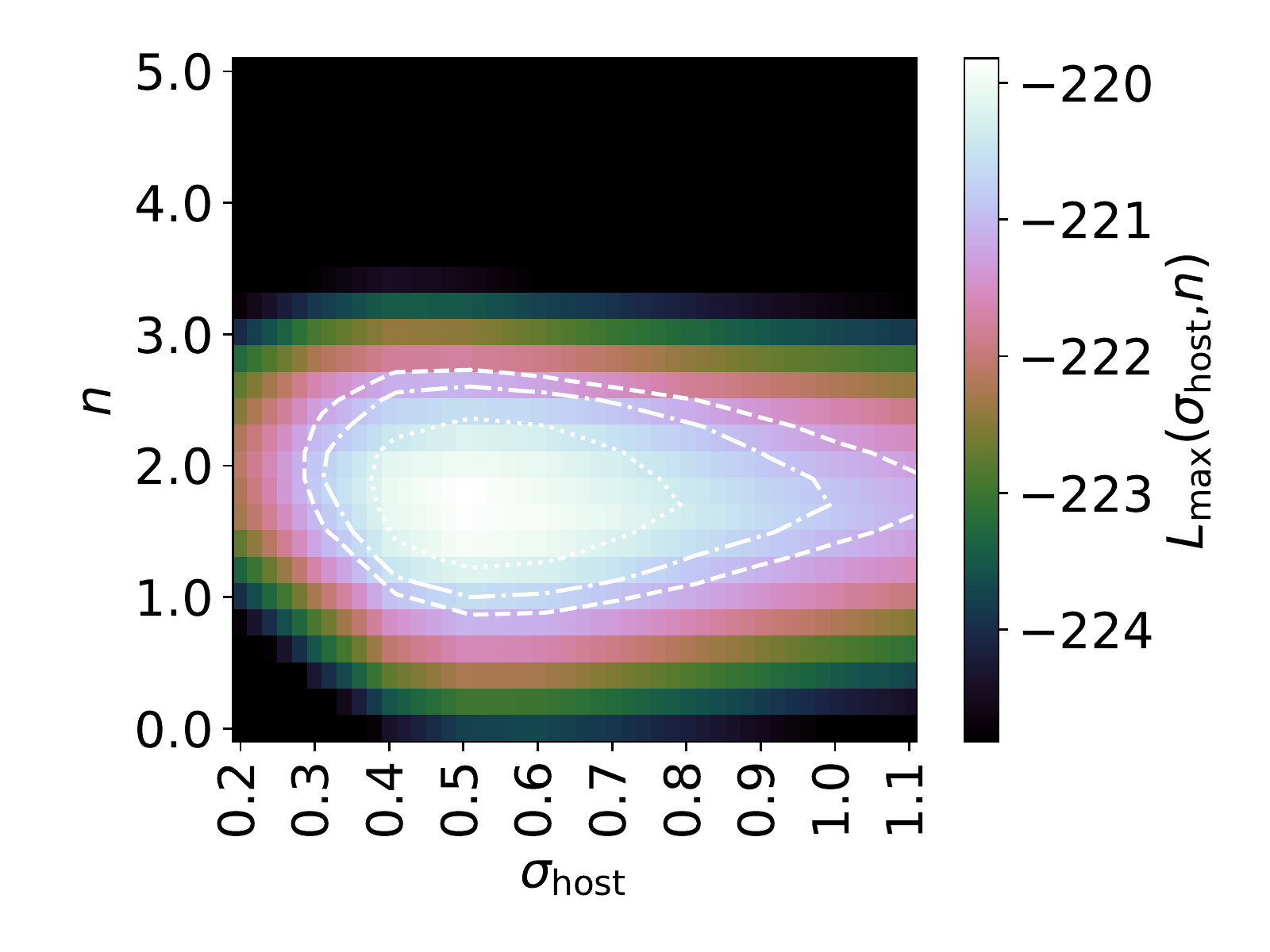}
        \includegraphics[width=0.32\textwidth]{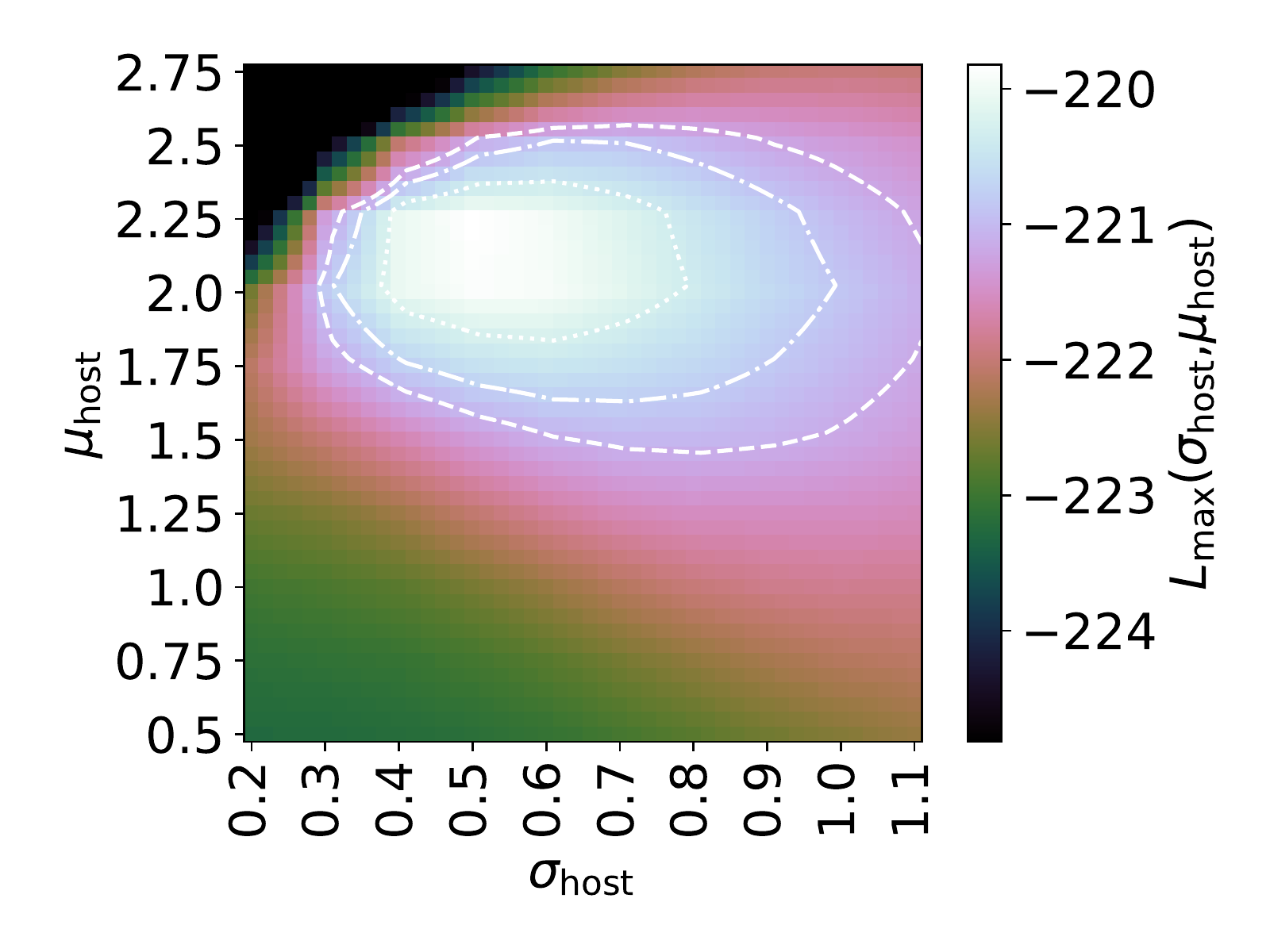}
    \caption{Two-parameter maximum likelihood results, showing 68\%, 90\%, and 95\% confidence intervals, calculated using Wilks' theorem and a $\chi^2_2$ distribution.}
    \label{fig:2dfigures}
\end{figure*}

Figure~\ref{fig:2dfigures} shows the correlation plots between all two-parameter combinations, excluding $\alpha$.

\begin{table*} 
\renewcommand{\arraystretch}{1.3}
    \centering
    \begin{tabular}{c|c c c c|c c c c}
     & \multicolumn{4}{|c|}{No prior} & \multicolumn{4}{|c|}{Prior on $\alpha$}\\
     Parameter & Best Fit & 68\% C.L.  & 90\% C.L.  & 95\% C.L. & Best Fit & 68\% C.L.  & 90\% C.L.  & 95\% C.L. \\
     \hline
$\log_{10} E_{\rm max}$ & 42.16 & $_{-0.38}^{+0.29}$  & $_{-0.56}^{+0.49}$  & $_{-0.56}^{+0.62}$  &  41.84 & $_{-0.18}^{+0.49}$  & $_{-0.24}^{+0.67}$  & $_{-0.24}^{+0.82}$ \\
$\alpha$ & -2.50 & N/A  & N/A  & N/A  &  -1.55 & $_{-0.21}^{+0.21}$  & $_{-0.33}^{+0.33}$  & $_{-0.41}^{+0.41}$ \\
$\gamma$ & -1.15 & $_{-0.13}^{+0.12}$  & $_{-0.19}^{+0.19}$  & $_{-0.22}^{+0.24}$  &  -1.16 & $_{-0.12}^{+0.11}$  & $_{-0.18}^{+0.20}$  & $_{-0.21}^{+0.26}$ \\
$n$ & 2.22 & $_{-0.35}^{+0.56}$  & $_{-0.96}^{+0.71}$  & $_{-1.16}^{+0.86}$  &  1.77 & $_{-0.45}^{+0.25}$  & $_{-0.66}^{+0.51}$  & $_{-0.81}^{+0.66}$ \\
$\mu_{\rm host}$ & 2.14 & $_{-0.25}^{+0.16}$  & $_{-0.41}^{+0.25}$  & $_{-0.55}^{+0.30}$  &  2.16 & $_{-0.23}^{+0.16}$  & $_{-0.39}^{+0.25}$  & $_{-0.50}^{+0.30}$ \\
$\sigma_{\rm host}$ & 0.54 & $_{-0.13}^{+0.17}$  & $_{-0.18}^{+0.35}$  & $_{-0.22}^{+0.47}$  &  0.51 & $_{-0.10}^{+0.15}$  & $_{-0.15}^{+0.30}$  & $_{-0.19}^{+0.41}$ \\

    \end{tabular}
    \caption{Confidence limits on single parameters, both with (left) and without (right) a prior on $\alpha$.}
    \label{tab:single_parameter_limits}
\end{table*}

In the main body of this work, we present parameter limits at 90\% C.L.\ obtained when using $\alpha$ as a prior. Table~\ref{tab:single_parameter_limits} presents parameter limits are different confidence levels, both with and without a prior on $\alpha$.
Furthermore, in Section~\ref{sec:modelling}, we note that the spectral index $\alpha$ can also be interpreted as a frequency-dependent rate. Our standard set of results use the spectral index interpretation. For completeness, in Table~\ref{tab:single_parameter_limits_rate}, we give the single-parameter limits under the rate interpretation.

\begin{table*} 
\renewcommand{\arraystretch}{1.3}
    \centering
    \begin{tabular}{c|c c c c|c c c c}
     & \multicolumn{4}{|c|}{No prior} & \multicolumn{4}{|c|}{Prior on $\alpha$}\\
     Parameter & Best Fit & 68\% C.L.  & 90\% C.L.  & 95\% C.L. & Best Fit & 68\% C.L.  & 90\% C.L.  & 95\% C.L. \\
     \hline
$\log_{10} E_{\rm max}$ & 41.93 & $_{-0.55}^{+0.22}$  & $_{-0.55}^{+0.48}$  & $_{-0.55}^{+0.63}$  &  41.86 & $_{-0.48}^{+0.28}$  & $_{-0.48}^{+0.51}$  & $_{-0.48}^{+0.69}$ \\
$\alpha$ & -2.50 & N/A  & N/A  & N/A  &  -1.50 & $_{-0.20}^{+0.21}$  & $_{-0.33}^{+0.35}$  & $_{-0.41}^{+0.41}$ \\
$\gamma$ & -1.19 & $_{-0.11}^{+0.17}$  & $_{-0.18}^{+0.26}$  & $_{-0.21}^{+0.30}$  &  -1.15 & $_{-0.16}^{+0.15}$  & $_{-0.22}^{+0.23}$  & $_{-0.25}^{+0.27}$ \\
$n$ & 1.97 & $_{-1.16}^{+0.25}$  & $_{-1.41}^{+0.51}$  & $_{-1.57}^{+0.61}$  &  1.26 & $_{-0.35}^{+0.51}$  & $_{-0.56}^{+0.76}$  & $_{-0.71}^{+0.86}$ \\
$\mu_x$ & 2.11 & $_{-0.23}^{+0.23}$  & $_{-0.36}^{+0.32}$  & $_{-0.50}^{+0.39}$  &  2.14 & $_{-0.23}^{+0.20}$  & $_{-0.36}^{+0.32}$  & $_{-0.48}^{+0.36}$ \\
$\sigma_x$ & 0.48 & $_{-0.10}^{+0.15}$  & $_{-0.15}^{+0.29}$  & $_{-0.18}^{+0.39}$  &  0.48 & $_{-0.10}^{+0.14}$  & $_{-0.15}^{+0.27}$  & $_{-0.18}^{+0.36}$ \\
    \end{tabular}
    \caption{Equivalent of Table~\ref{tab:single_parameter_limits}, calculated assuming the rate interpretation of $\alpha$ (see Section~\ref{sec:modelling}).}
    \label{tab:single_parameter_limits_rate}
\end{table*}

\bsp	
\label{lastpage}
\end{document}